\documentclass[conference]{IEEEtran}
\IEEEoverridecommandlockouts
\usepackage{booktabs} 
\usepackage{times} 
\usepackage{color} 
\usepackage{balance} 
\usepackage{url} 
\usepackage[all]{nowidow}
\usepackage{tabularx}
\usepackage{siunitx} 
\usepackage{tabularx,subfigure,multirow, graphicx}
\usepackage{epsfig}
\usepackage{epstopdf} 
\usepackage{enumitem}
\usepackage{caption}
\usepackage[a-2b]{pdfx}

\usepackage[pdfa]{hyperref} 

\usepackage{amsthm,amsmath}  

\newcolumntype{L}[1]{>{\raggedright\arraybackslash}p{#1}}
\newcolumntype{C}[1]{>{\centering\arraybackslash}p{#1}}
\newcolumntype{R}[1]{>{\raggedleft\arraybackslash}p{#1}}

\usepackage[linesnumbered,ruled,vlined]{algorithm2e}
\SetKwRepeat{Do}{do}{while}
\SetCommentSty{mycommfont}

\long\def\comment#1{}

\DeclareMathOperator*{\argmin}{arg\,min}

\newcommand{\nop}[1]{}

\newtheorem{lemma}{\bf Lemma}[section]

\theoremstyle{remark}

\theoremstyle{definition}
\newtheorem{definition}{\bf Definition}

\begin{document}
	
\title{Approximate Nearest Neighbor Search of Large Scale Vectors on Distributed Storage}

\author{
	Kun Yu{\small$~^{1}$},
	 Jiabao Jin{\small$~^{2}$},
	 Xiaoyao Zhong{\small$~^{2}$}, 
	 Peng Cheng{\small$~^{3}$}, 
	 Lei Chen{\small$~^{4,5}$},\\
	 Zhitao Shen{\small$~^{3}$},
	 Jingkuan Song{\small$~^{3}$}, 
	 Hengtao Shen{\small$~^{3}$},
	 Xuemin Lin{\small$~^{6}$}\\
	\fontsize{10}{10}\itshape
	$~^{1}$ECNU, Shanghai, China;
	$~^{2}$Ant Group, Shanghai, China;
	$~^{3}$Tongji University, Shanghai, China;\\
	$~^{4}$HKUST (GZ), Guangzhou, China;
	$~^{5}$HKUST, Hong Kong SAR, China;
	$~^{5}$SJTU, Shanghai, China\\
	\fontsize{9}{9}\upshape
	kunyu@stu.ecnu.edu.cn; jinjiabao.jjb@antgroup.com; zhongxiaoyao.zxy@antgroup.com; cspcheng@tongji.edu.cn;\\ leichen@cse.ust.hk; zhitao.szt@antgroup.com; jingkuan.song@gmail.com; shenhengtao@hotmail.com; xuemin.lin@gmail.com
}

\maketitle
\begin{abstract}
	Approximate Nearest Neighbor Search (ANNS) in high-dimensional space is an essential operator in 
	many online services, such as  information retrieval and recommendation. 
	Indices constructed by the state-of-the-art ANNS algorithms must be stored in
	single machine's memory or disk for high recall rate and throughput, 
	suffering from substantial storage cost, constraint of limited scale and single point of failure.
	While distributed storage can provide a cost-effective and robust solution,
	there is no efficient and effective algorithms for indexing vectors in distributed storage scenarios.
	In this paper, we present a new graph-cluster hybrid indexing and search system
	which supports \underline{D}istributed \underline{S}torage 
	\underline{A}pproximate \underline{N}earest \underline{N}eighbor Search, called DSANN.
	DSANN can efficiently index, store, search billion-scale vector database in distributed storage
	and guarantee the high availability of index service.
	DSANN employs the concurrent index construction method to significantly reduces the complexity of index building. 
	Then, DSANN applies Point Aggregation Graph to leverage the structural information of graph to aggregate similar vectors,
	optimizing storage efficiency and improving query throughput via asynchronous I/O in distributed storage.
	Through extensive experiments, we demonstrate DSANN can efficiently and effectively index, store and search
	large-scale vector datasets in distributed storage scenarios.
\end{abstract}

\section{Introduction}
\label{sec:Introduction}

With the continuous expansion and development of content community platforms, such as YouTube~\cite{youtube} and TikTok~\cite{tiktok}, 
the growth rate of unstructured data (e.g., text, images, videos, and audio) far exceeds that of structured data.
Since vectors are the primary semantic representation  form of unstructured data~\cite{DBLP:journals/corr/abs-1301-3781}, 
approximate nearest neighbor search (ANNS)~\cite{DBLP:conf/soda/AryaM93, DBLP:journals/jacm/AryaMNSW98} on vectors has become an infrastructure in information retrieval field 
and is widely used in AI-related fields such as computer vision and recommendation system~\cite{DBLP:conf/cvpr/WangWZTGL12, DBLP:conf/kdd/LiLJLYZWM21, DBLP:conf/ifip12/SuchalN10}. 
Especially with the rise of large language models (LLMs)~\cite{DBLP:journals/corr/abs-2305-04359}, 
ANNS has been applied in retrieval-augmented generation (RAG) 
to improve the accuracy and reliability of LLMs~\cite{DBLP:journals/corr/abs-2312-10997}.

However, the exponential growth of data presents unprecedented challenges for efficient vector retrieval. 
For instance, YouTube experiences an average upload rate of over 500 hours of video content per minute, 
and Alibaba processes approximately 500 PB of unstructured data during shopping festivals~\cite{youtube, DBLP:journals/pvldb/WeiWWLZ0C20}. 
This surge in data scale has exposed the limitations of existing methods in terms of cost-effectiveness 
when handling massive vector retrieval tasks. 
Besides, many online vector retrieval tasks often require multiple replicas to ensure high availability of the service, 
which significantly increases storage costs on dealing with larger-scale datasets.
Consequently, there is \emph{an urgent need for
	cost-effective solutions} to address these emerging challenges in large-scale data processing and retrieval.

\emph{Distributed storage solutions such as Distributed File System (DFS) and Object Storage (OS) 
	offer cost-effective and highly available storage}~\cite{bs, s3, ebs} compared to memory and local disk storage~\cite{DBLP:journals/tocs/ChangDGHWBCFG08, DBLP:conf/mss/ShvachkoKRC10, factor2005object}. 
Moreover, in a disaggregated storage architecture, compute and storage are decoupled, 
allowing system designers to scale each resource independently. 
This not only reduces the overall risk of single point failure 
but also offers a cost advantage in scenarios with variable or lower request volumes, 
where one can flexibly provision fewer compute nodes while relying on shared storage.
Such an architecture eliminates redundant storage of index files across replicas, significantly reducing storage overhead. 
Unlike local disk-based architectures, where each replica maintains a full index copy, 
distributed storage enables multiple computing nodes to access a shared index, 
minimizing redundancy and lowering storage costs. 
Additionally, distributed storage accelerates failover recovery by removing the need to reload index files onto local disks, 
ensuring faster recovery and higher availability.
Unfortunately, \emph{distributed storage typically has higher I/O overheads}
compared to memory and local disk storage. 
The read latency of distributed storage can be significantly slower,
often by an order of magnitude or more, than that of local disk or memory access.

\begin{table}[ht!]
	\caption{Pangu Storage System~\cite{bs, DBLP:conf/fast/LiXWSWYDZHZWLCW23} vs. Local SSD}
	\label{tab:comparison}
	\centering
	\begin{tabular}{l|l|l}
		\hline
		\textbf{Feature}                    & \textbf{Pangu Storage System}                    & \textbf{Local SSD}       \\ \hline\hline
		{Type}                       & Distributed Storage                        & SSD                       \\ \hline
		{Durability}                 & 99.9999999\%                        & Single Point of Failure                  \\ \hline
		{Latency}                    & 0.1-10 ms                               & 1-100 us                       \\ \hline
		{Scalability}                & Unlimited                             & Limited                      \\ \hline
	\end{tabular}
\end{table}

Memory-disk hybrid ANNS solutions can be broadly classified into two main categories: 
\emph{graph-based}~\cite{jayaram2019diskann} 
and \emph{cluster-based}~\cite{DBLP:conf/icassp/JegouTDA11, DBLP:conf/nips/ChenZWLLLYW21, DBLP:conf/cvpr/YandexL16} methods. 
Among them, DiskANN and SPANN represent the state-of-the-art techniques 
in graph-based and cluster-based methods, respectively.
DiskANN stores Product Quantization (PQ)~\cite{DBLP:journals/pami/JegouDS11, DBLP:conf/cvpr/KalantidisA14} compressed vectors in memory 
while keeping the graph and full-precision vectors on disk. 
SPANN uses hierarchical balanced clustering (HBC) to generate a large number of balanced partitions, 
storing only the cluster centroids in memory while putting partition lists on disk.

\begin{figure}[ht!]
	\centering\vspace{-2ex}
	\subfigure[next candidate selection]{\includegraphics[width=0.49\linewidth]{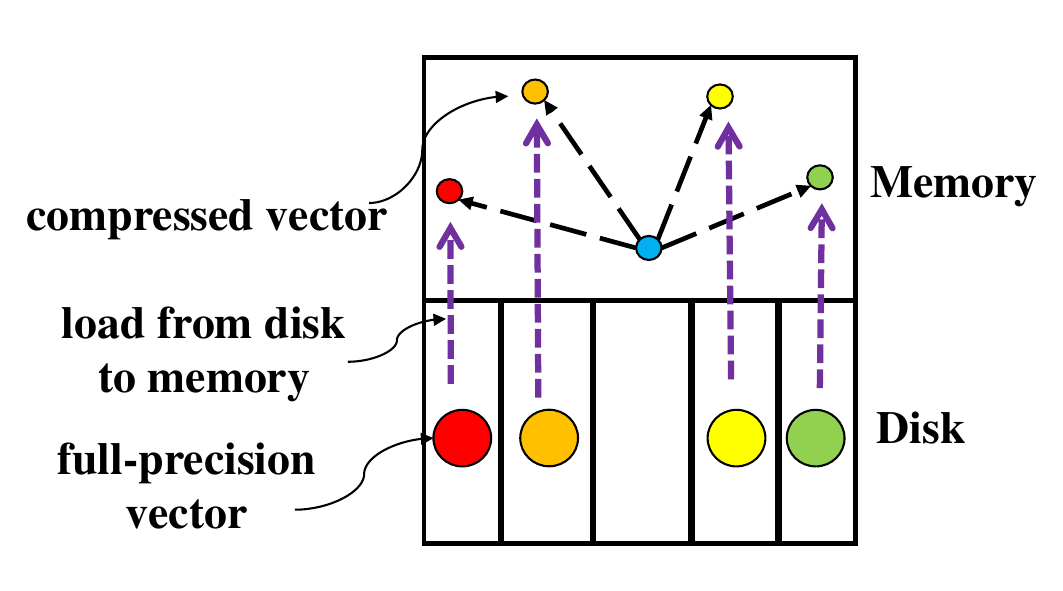}\label{fig:exist_problem_a}} 
	\subfigure[boundary problem]{\includegraphics[width=0.49\linewidth]{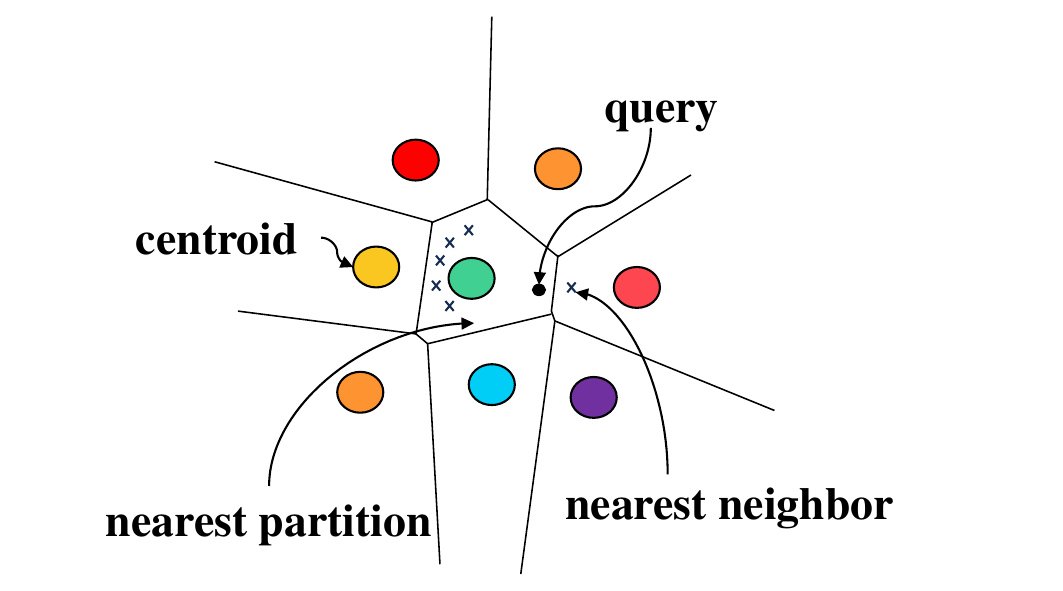}\label{fig:exist_problem_b}}
	\caption{Existing problems in memory-disk hybrid ANNS solutions}
	\label{fig:exist_problem}
\end{figure}

Although memory-disk hybrid ANNS solutions are relatively mature, 
they struggle to scale effectively to distributed storage.
Candidate reordering process in DiskANN requires loading full-precision vectors from 
secondary storage into memory, resulting in significant delays as shown in Figure~\ref{fig:exist_problem_a},
where circles of different colors represent distinct vectors. 
The partition-based approach used by SPANN often suffers from the curse of dimensionality and boundary issues, 
which leads to lower accuracy, as illustrated in Figure~\ref{fig:exist_problem_b},
where circles of different colors denote partition centroids.

Generally, there are two challenges that must be addressed 
when extending ANNS to distributed storage.

\noindent\underline{Challenge I:} \textit{How to deal with high I/O latency of ANNS on distributed storage?}
While distributed storage presents several advantages for ANNS, 
it introduces higher I/O latency compared to memory and disk, as summarized in Table~\ref{tab:comparison}.
DiskANN requires loading full-precision vectors into memory during the candidate reranking stage
before proceeding with subsequent search processes, leading to substantial delays.
Similarly, SPANN needs I/O operations to fetch partition lists into memory before the fine-grained search, which incurs high I/O cost.
This issue becomes more pronounced in huge vector retrieval scenarios, 
where ANNS is expected to recall thousands of vectors.
In such cases, DiskANN demands more I/O operations due to an increased number of routing hops,
while SPANN must retrieve additional vectors as it explores more partitions.
Particularly, I/O operations block the subsequent calculations of DiskANN, causing higher latency.
Therefore, it is desirable to design algorithms that reasonably increase the search fan-out and support asynchronous I/O.

\noindent\underline{Challenge II:} \textit{How to accelerate the index construction for very large-scale datasets?}
Despite the scalability of distributed storage in managing large-scale datasets,
efficient index construction for billion-scale datasets remains a significant challenge.
Both DiskANN and SPANN require several days to build the index,
which is unacceptable for online services~\cite{DBLP:conf/sosp/XuLLXCZLYYYCY23}.
Specifically, DiskANN incurs significant construction overhead due to the need 
for redundant vector replication across multiple partitions when building the proximity graph. 
Similarly, SPANN aims to achieve balanced partitions through multiple iterations and refinements, 
resulting in substantial computational and storage costs.
However, with the availability of many cost-effective machines within technology companies,
it is worthwhile to design algorithms that support concurrent index construction,
which can also be easily extended to multiple machines for parallel index construction.

In this paper, we propose DSANN, a novel graph-cluster hybrid index designed for large-scale ANNS.
DSANN efficiently constructs the index by leveraging the structural information of proximity graph
to aggregate multiple similar points (denoted as residual points) into a single representative point (denoted as an aggregation point) in the graph.
To enable efficient storage and retrieval, 
DSANN maintains aggregation points in memory while organizing residual points as partition lists on secondary storage.
Moreover, DSANN employs a partition-construction-merge strategy to accelerate the index construction,
which naturally extends to parallel construction across multiple machines.
Once the index is built, 
DSANN optimizes the search process by balancing I/O and computation overheads
through an asynchronous query execution mechanism,
ensuring efficient and scalable search operations.

\noindent \textbf{Contributions.} It is worthwhile to highlight our contributions as follows:
\begin{itemize}[leftmargin=10pt, topsep=1pt]
	\item{} We are the first to propose an ANNS solution on distributed storage, 
	which can significantly reduce storage cost while supporting larger-scale vectors. 
	The disaggregated architecture not only avoids redundant index replication but also adapts to varying workloads, 
	offering cost-efficient deployment even under lower request volume scenarios.
	Furthermore, we design a new index structure that balances the I/O and computation costs 
	in an asynchronous manner during the search process in Section~\ref{sec:framework}.
	\item{} We present a concurrent index construction algorithm to accelerate the process of index construction,
	which can be easily extended to multiple machines for parallel construction.
	Through complexity analysis and extensive experiments, 
	we demonstrate that concurrent index construction is more efficient than traditional methods in Section~\ref{sec:construction}.
	\item{} Through a comprehensive experimental study, 
	DSANN effectively meets the vector retrieval needs across both large and small scales, 
	demonstrating its wide applicability and efficiency in different application scenarios. 
	Our experiments show that DSANN can significantly reduce latency on distributed storage
	when handling large-scale vector retrieval while maintaining low computation cost during small-scale vector retrieval in Section~\ref{sec:experimental}.
\end{itemize}
\section{PRELIMINARIES}
\label{sec:preliminary}
In this section, we present relevant preliminaries and 
introduce state-of-the-art ANNS methods which can be extended to distributed scenarios.
Frequently used notations are summarized in Table~\ref{tab:notation}.

\begin{table}[ht]
	\centering
		\caption{Symbols and Descriptions}
		\label{tab:notation}
		\begin{tabular}{l|l}
			\hline
			{\bf Notation}          & {\bf Description} \\ 
			\hline \hline
			$E^d$                   & d-dimensional Euclidean space\\
			$\mathcal{X}$           & a finite dataset of $n$ vector $\vec{x}$\\
			$\vec{x}_q$             & the query vector\\
			$\delta(\cdot, \cdot)$  & the squared Euclidean distance between two vectors\\
			$G(V,E)$                & a proximity graph $G$ with vertices $V$ and edges $E$\\
			$NN(\vec{x})$           & neighbors of $\vec{x}$\\
			\hline
			\hline
		\end{tabular}
\end{table}

\begin{figure*}[ht!]
	\centering
	\includegraphics[width=0.90\linewidth]{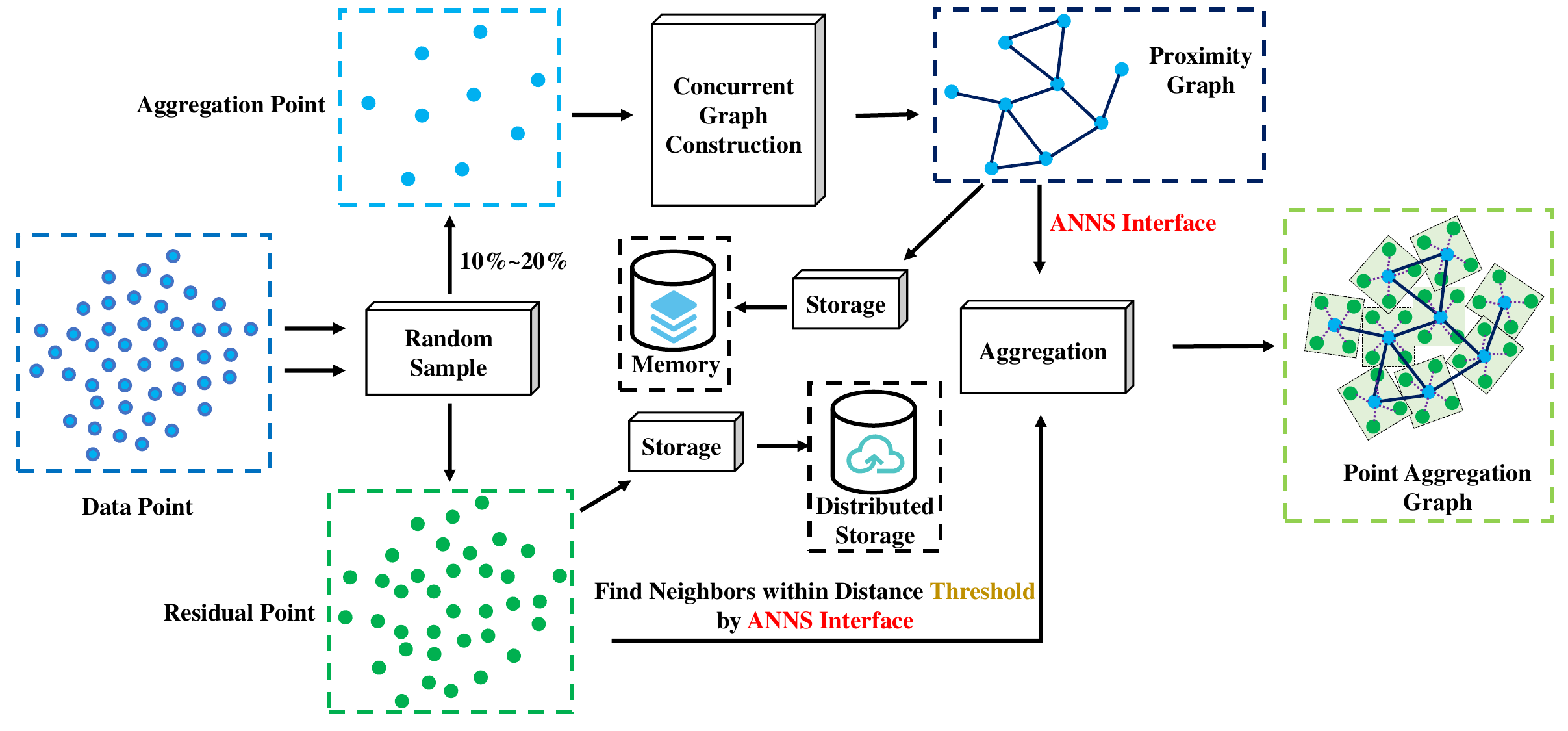}
	\caption{Illustration of the graph construction}
	\label{fig:pag_construction}
\end{figure*}

\subsection{Problem Setting}
Various problems in information retrieval and management of high-dimensional vector data 
can be abstracted as the nearest neighbor search problem in high-dimensional space.
Nearest Neighbor Search (NNS) aims to find the closest vector from the dataset for a given query.
NNS can be defined as follows:
\begin{definition}
	\label{def:nns}
	\textbf{Nearest Neighbor Search}. Given a finite vector dataset $\mathcal{X}$ of $n$ vectors in space $E^d$ and a query vector $\vec{x}_q$.
	NNS aims at efficiently obtaining a vector $\vec{x}_p \in \mathcal{X}$ that is closest to $\vec{x}_q$.
\end{definition}

Exact solutions for NNS can not be applied in large scale dataset due to 
the expensive computation cost and high query latency.
Therefore, Approximate Nearest Neighbor Search (ANNS) \cite{DBLP:conf/soda/AryaM93} has been proposed to 
trade a little loss in accuracy for much higher efficiency
and can be defined as follows:
\begin{definition}
	\label{def:anns}
	\textbf{Approximate Nearest Neighbor Search}. Given a finite vector dataset $\mathcal{X}$ of $n$ vectors in space $E^d$,
	a query vector $\vec{x}_q$ and a parameter $\epsilon > 0$.
	ANNS aims at efficiently obtaining a vector $\vec{x}_p \in \mathcal{X}$ that is close to $\vec{x}_q$, 
	satisfying the condition: $\delta(\vec{x}_p, \vec{x}_q) \le (1 + \epsilon)\delta(\vec{x}_r, \vec{x}_q)$,
	where $\vec{x}_r$ is the closest vector to $\vec{x}_q$ in $\mathcal{X}$ 
\end{definition}

This problem can naturally generalize to $K$ Approximate Nearest Neighbor Search (KNNS) 
where we need to return the closest $k > 1$ vectors to $\vec{x}_q$.
For the convenience of evaluating accuracy, we usually use \emph{recall@$k$} instead of the $\epsilon$ in practice.
Given a query $\vec{x}_q$,
$\mathcal{R}$ is the result set of $k$ vectors obtained by ANNS,
and $\widetilde{\mathcal{R}}$ is the accurate $k$ nearest neighbor set of $\vec{x}_q$.
Then we can define \emph{recall@$k$} as follows:
\begin{equation}
	\label{eq:recall}
	Recall@k=\frac{|\mathcal{R} \cap \widetilde{\mathcal{R}}|}{|\widetilde{\mathcal{R}}|}=\frac{|\mathcal{R} \cap \widetilde{\mathcal{R}}|}{k}
\end{equation}
It is worth mentioning that coarse-grained sort usually requires obtaining
huge vectors from the dataset, which is widely used in recommendation system.
Therefore, we should not only consider the small $k$ scenarios,
but also scale the $k$ to ten thousands.

Even though current mainstream ANNS methods have done well on memory and disk of a single machine,
companies often need to separate computation and storage for the consideration of cost, 
such as storing large-scale data in more cost-effective storage like DFS or OS.
Nevertheless, cost-effective storage can lead to increased network I/O latency, 
presenting a significant challenge.

\section{DSANN FRAMEWORK}
\label{sec:framework}

We propose a novel ANNS methods DSANN,
optimized for cost-effective storage and huge vector retrieval scenarios.
DSANN combines the advantages of cluster-based and graph-based ANNS approaches.
We introduce the motivation behind the idea in Section~\ref{sec:intuition}
and provide a brief description of DSANN framework in Section~\ref{sec:overview}

\subsection{Motivation} \label{sec:intuition}
\begin{figure}[ht!]
	\centering\vspace{-2ex}
	\includegraphics[width=0.98\linewidth]{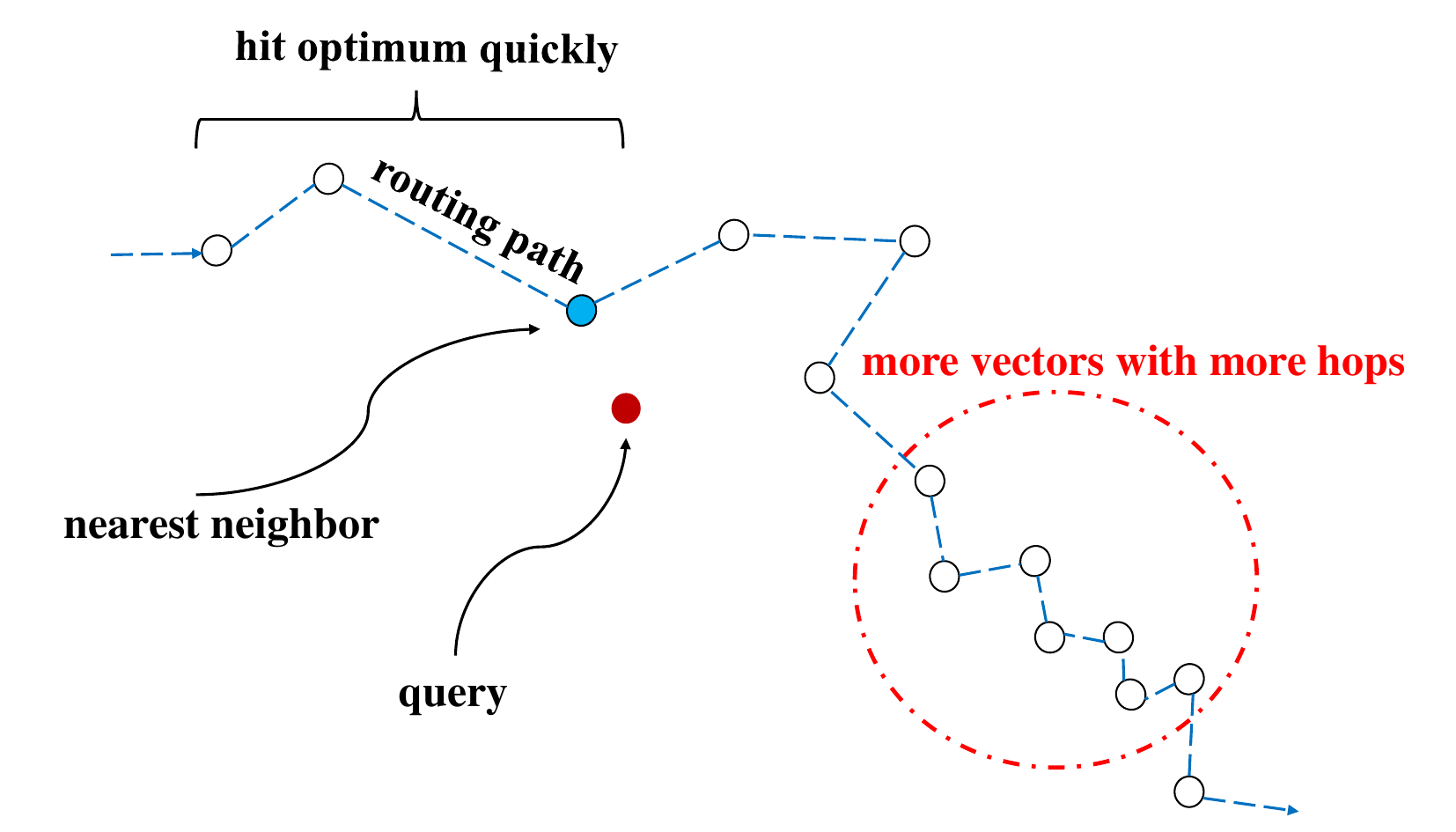}
	\caption{Recall more vectors with longer routing path}
	\label{fig:routing_path}
\end{figure}

Current state-of-the-art ANNS methods benefit from their index structure for supporting the hybrid-storage scenarios.
Graph-based ANNS methods reduce the times of I/O by optimizing the proximity graph structure for shorter search path~\cite{jayaram2019diskann}
while cluster-based ANNS methods seek to obtain better partitions for fewer partition probes to lower the I/O cost~\cite{DBLP:conf/nips/ChenZWLLLYW21}.
Nevertheless, distributed storage scenarios will introduce higher I/O latency and 
the mainstream ANNS methods such as DiskANN have to stall until the I/O completion.
An effective approach to reducing I/O overhead is asynchronous I/O, 
which decouples computation from data retrieval, enabling their concurrent execution.
Besides, graph-based ANNS methods usually need to significantly increase the size of candidates for obtaining huge vectors,
leading to longer routing path and higher latency as shown in Figure~\ref{fig:routing_path}. 
On the contrary, cluster-based ANNS methods are well-suited for huge vector retrieval
because it can obtain more results by exploring more partitions.
However, cluster-based methods often suffer from the boundary problem which may lead to much more useless exploration.
More importantly, previous methods require several days to construct the index on a single powerful and expensive server,
despite there are numerous less powerful yet economical servers in companies.
Therefore, employing multiple less powerful machines for parallel index construction emerges 
as a promising strategy of cost-effective ANNS.

From our perspective, we can improve the ANNS performance of distributed storage scenarios
by combining the advantages of graph-based and cluster-based ANNS methods.
Graph-based traversal makes asynchronous I/O and computation possible
and partition exploration is well adapted to huge vector retrieval because of large fan-out.
In summary, we aim to design a new ANNS index with high performance from the following aspects.
\emph{
	(1) making I/O and computation decouple and balance,
	(2) supporting huge vector retrieval,
	(3) working on distributed-storage scenarios.
	(4) employing numerous cost-effective machines for parallel index construction
}
Point (1) can be achieved by the traversal of graph 
while fetching vectors from distributed storage via network I/O.
For Point (2)-(3), the idea of clustering is similar to compression
and partition exploration is well-adapted to huge vector retrieval.
Regarding Point (4), we can distribute the data across multiple machines to construct localized proximity graphs, 
followed by subsequent merging.
Below we propose a new graph-based structure called Point Aggregation Graph,
which aggregates multiple close vectors into a vertex of the proximity graph.
Thanks to the distribution of the index structure,
DSANN can achieve better performance in cost-effective storage
and huge vector retrieval scenarios.

\subsection{Overview} \label{sec:overview}

\begin{figure*}[ht!]
	\centering\vspace{-2ex}
	\includegraphics[width=0.90\linewidth]{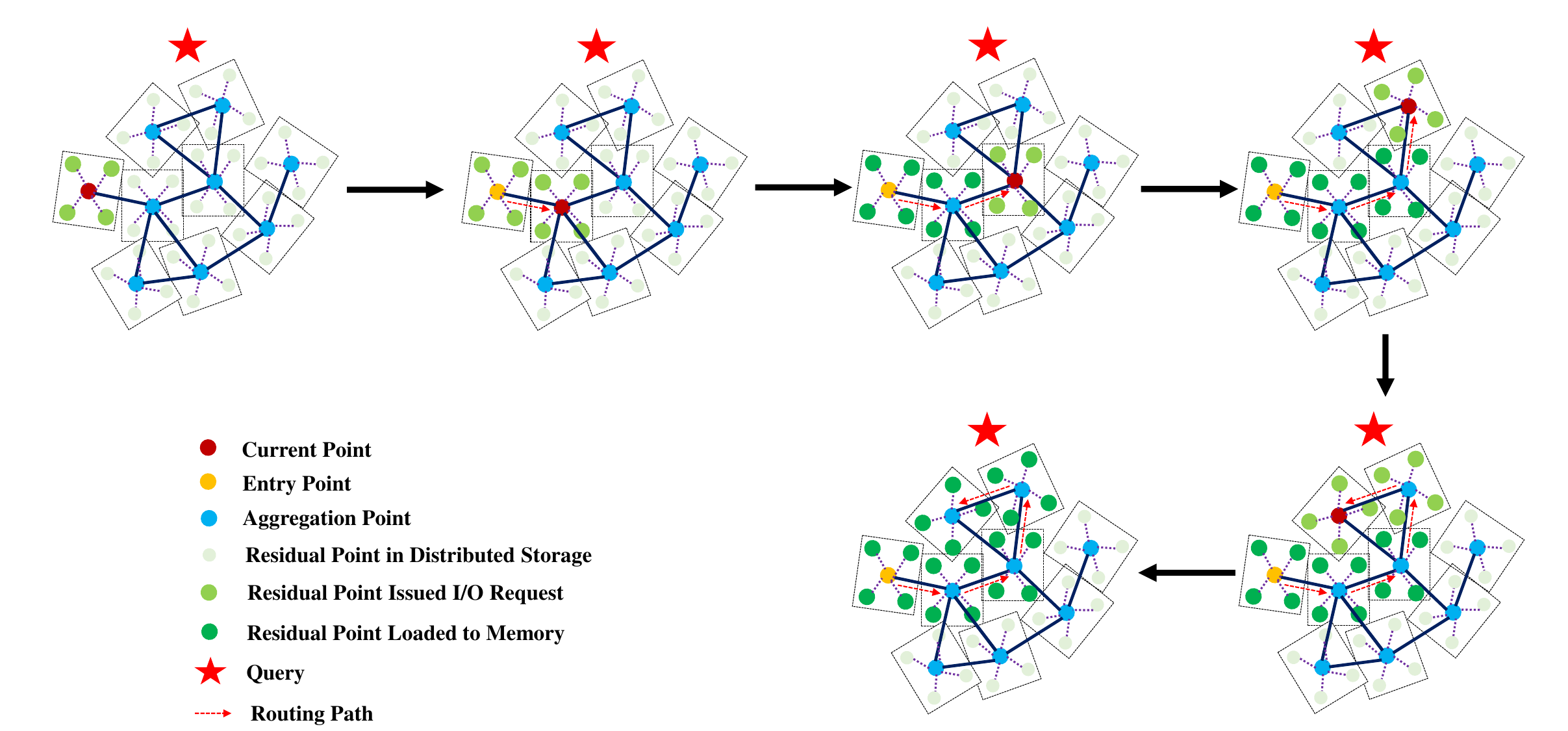}
	\caption{Illustration of the search on graph}
	\label{fig:pag_search}
\end{figure*}

The overall pipeline of Point Aggregation Graph (PAG) construction is illustrated in Figure~\ref{fig:pag_construction}.
DSANN samples a small portion (10\%-25\%) of data as aggregation points for the concurrent construction of the proximity graph. 
The rest data points called residual points are aggregated to the nearest aggregation points within a distance limit
by the search Algorithm~\ref{algo:greedy_search} on the graph.
In addition, to improve the performance of the index, 
we leverage the structural information of the graph to choose multiple redundant aggregation points for residual points.
And then DSANN keeps aggregation points and the proximity graph in memory 
while storing the residual points in the distributed storage for the separation of computation and storage.
In Section~\ref{sec:construction}, we will provide detailed analysis of PAG and optimize it for lower query latency.
Figure~\ref{fig:pag_search} illustrate the process of search on the index.
DSANN employs Algorithm~\ref{algo:greedy_search} to traverse the proximity graph 
while issuing I/O requests to fetch relevant residual points 
based on the distance between query and aggregation points
from the distributed storage for full scan.
In order to avoid unnecessary exploration, 
DSANN employs early stop that adaptively terminates graph traversal for each query.
The core idea of decoupling the computation and I/O involves 
distributing the relevant residual points across the routing path. 
Further details about search will be discussed in Section~\ref{sec:search}

\begin{algorithm}[t]
	\DontPrintSemicolon
	\KwIn{Proximity graph $G$ with entry point $s$, query point $\vec{x}_q$, search parameter $L$, $K$}
	\KwOut{Approximate $K$ nearest neighbors of $\vec{x}_q$}
	
	initialize a candidate set $C$ as a maximum-heap with size $L$ \;
	$C \leftarrow C \cup \{ s \}$ \;
	
	\While{$C$ has unexpanded nodes}{
		$c \leftarrow$ closest unexpanded node in $C$ \;
		calculate the distance between query $\vec{x}_q$ and each
		neighbor of $c$ \;
		insert the neighbors into $C$
	}
	
	\Return{$K$ nearest points in $C$}\;
	\caption{Greedy Search on Graph}
	\label{algo:greedy_search}
\end{algorithm}

\section{Index Construction}
\label{sec:construction}
\subsection{Point Aggregation Graph}
Graph-based \cite{DBLP:journals/pvldb/WangXY021} and cluster-based \cite{DBLP:conf/cvpr/BabenkoL12, DBLP:conf/eccv/BaranchukBM18} ANNS methods 
benefit from their respective index structure.
Graph-based methods provide high accuracy while cluster-based methods can be viewed as a form of compression.
It is intuitive to combines the advantages of them to design a high-quality index structure
that is tailored for distributed storage.

Consider a proximity graph $G(V, E)$ constructed on a dataset $\mathcal{X}$ 
consisting of billions of data points or even more. 
While ANNS on $G$ using Algorithm~\ref{algo:greedy_search} can achieve state-of-the-art performance, 
it requires substantial memory to keep the index for online search.
Following the cluster-based methodology, 
we can aggregate the close vectors into a vertex of the graph stored in the memory, 
while the remaining points are organized as partition lists in the distributed storage.
We call this index structure the Point Aggregation Graph.
Before presenting our proposal, we first provide several essential definitions as follows:
\begin{definition}
	\label{def:pg}
	\textbf{Proximity Graph (PG)}. Given a finite vector dataset $\mathcal{X}$ of $n$ vectors in space $E^d$
	and a distance threshold $\theta > 0$. 
	A proximity graph $G(V, E)$ w.r.t $\theta$ on $\mathcal{X}$ is defined as follows:
	(1) For each vector $\vec{x}_i \in \mathcal{X}$, there exists a corresponding vertex $v_i \in V$.
	(2) For any two vertices $v_i$, $v_j \in V$, if an edge $v_iv_j \in E$ exists, 
	then it holds that the distance $\delta(\vec{x}_i, \vec{x}_j) \leq \theta$.   
\end{definition}

\begin{figure}[ht!]
	\centering\vspace{-2ex}
	\subfigure[proximity graph]{\includegraphics[width=0.49\linewidth]{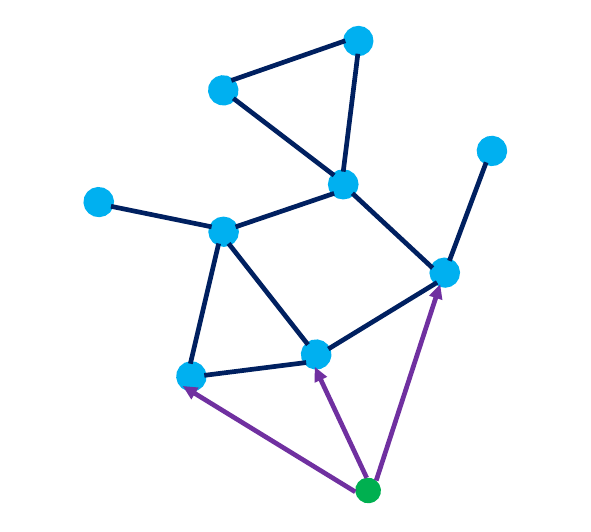}\label{fig:difference_a}} 
	\subfigure[point aggregation graph]{\includegraphics[width=0.49\linewidth]{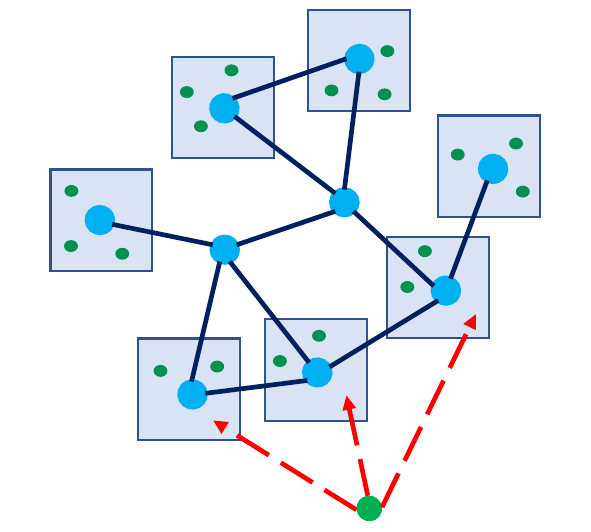}\label{fig:difference_b}}
	\caption{Difference between PG and PAG}
	\label{fig:difference}
\end{figure}

\begin{definition}
	\label{def:pag}
	\textbf{Point Aggregation Graph (PAG)}. Given a finite vector dataset $\mathcal{X}$ of $n$ vectors in space $E^d$,
	a PG $G_p(V^a, E^a)$ on aggregation point set $\mathcal{X}^a \subseteq \mathcal{X}$, 
	residual point set is $\mathcal{X}^r = \mathcal{X} \backslash \mathcal{X}^a$ and an aggregation radius $r$.
	A point aggregation graph $G(V, E)$ is defined as follows:
	(1) For each vector $\vec{x}^r_i \in \mathcal{X}^r$, there exists a corresponding vertex $v^r_i \in V^r = V \backslash V^a$.
	(2) For any two vertices $v^a_i \in V^a$, $v^r_j \in V^r$, if an edge $v^a_iv^r_j$ exists,
	then it holds that the distance $\delta(\vec{x}^a_i, \vec{x}^r_j) \leq r$.
	(3) For each vertex $v^r_i \in V^r$, there exist a unique edge $v^a_jv^r_i \in E \backslash E^a$, where $v^a_j \in V^a$. 
\end{definition}
The structure of PAG is illustrated in Figure~\ref{fig:pag_construction}.
PAG categories the data points into two types: \emph{aggregation points} and \emph{residual points}.
Aggregation points are used to construct a PG and 
each aggregation point is surrounded by several residual points, which together form a fully connected graph.
It is noteworthy that every fully connected graph can be logically viewed as a partition.
The key difference between PAG and PG is that 
PAG represents the proximal relationship through partition containment,
as shown in Figure~\ref{fig:difference_b}
while PG uses edge links, as shown in Figure~\ref{fig:difference_a}.

Although the structure of PAG is relatively simple,
constructing such graph incurs significant computational costs, as each residual point needs linear scan 
to identify its corresponding aggregation point.
Considering the much faster ANNS interface provided by the PG,
each residual point can efficiently find its nearest neighbor as aggregation point by graph traversal.
This approach constructs an approximate PAG, which we refer to naive PAG 
and the construct process is summarized in Algorithm \ref{algo:naive_pag_build}.

Specifically, during the construction of naive PAG, 
DSANN initially samples a proportion $0 < p < 1$ of the data points from the entire dataset $\mathcal{X}$ as 
aggregation points $\mathcal{X}^a$ to construct a PG $G(V^a, E^a)$, such as HNSW or Vamana \cite{DBLP:journals/pami/MalkovY20, jayaram2019diskann}.
Notably, the sample rate $p$ can be conceptually viewed as the compression rate.
For the residual data points $\mathcal{X}^r = \mathcal{X} \backslash \mathcal{X}^a$, 
we employ the ANNS interface provided by the graph $G(V^a, E^a)$
to identify the nearest neighbor $\vec{x}^a_i \in \mathcal{X}^a$ for each point $\vec{x}^r_i \in \mathcal{X}^r$.
Subsequently, each point $\vec{x}^r_i$ is assigned to the partition associated with $\vec{x}^a_i$. 
Following these assignments, DSANN keeps the PG $G$ and aggregation points $\mathcal{X}^a$ in memory, 
while putting the residual points $\mathcal{X}^r$ in distributed storage.

We assert that the time complexity of the PAG construction algorithm is 
significantly lower than that of DiskANN and SPANN.
The time complexity of Algorithm~\ref{algo:naive_pag_build} can be approximated as follows:
\begin{equation}
	O(pn \log{pn} + (1 - p)n \log{pn}) = O(n \log{pn})
\end{equation}

The first term corresponds to the construction of a proximity graph over $pn$ points,
which incurs a complexity of $O(pn \log pn)$.
The second term accounts for the search operation on the proximity graph with $pn$ points,
which has a time complexity of $O(\log pn)$.
In comparison, the construction complexity of DiskANN is $O(n \log n)$,
and SPANN exhibits a complexity exceeding $O(n \log n)$, as reported in their original work.
Notably, since $p < 1$, it follows that $\log pn < \log n$, 
implying that the PAG construction algorithm achieves a lower complexity.
Experimental results presented in Section~\ref{sec:experimental} empirically validate this theoretical advantage.
The claim is summarized as the following lemma.
\begin{lemma}
	The time complexity of the PAG construction algorithm is $O(n \log{pn})$, which is strictly lower than the construction complexities of DiskANN and SPANN.
\end{lemma}



\begin{algorithm}[t]
	\DontPrintSemicolon
	\KwIn{A vector dataset $\mathcal{X}$ of $n$ vectors}
	\KwOut{A naive PAG $G(V, E)$}
	
	$\mathcal{X}^a \leftarrow$ random sample from $\mathcal{X} $ \;
	\tcp*[l]{keep the fully connected graph as partition}
	$\mathcal{L} \leftarrow$ partition lists of $\left\lvert \mathcal{X}^a \right\rvert$ aggregation points \;
	$G_p(V^a, E^a) \leftarrow$ PG on $\mathcal{X}^a$ \;
	
	\ForEach{$\vec{x}^r_i \in \mathcal{X}^r = \mathcal{X} \backslash \mathcal{X}^a$}{
		retrieve the nearest neighbor $\vec{x}^a_i$ of $\vec{x}^r_i$ by the greedy search on $G_p(V^a, E^a)$\;
		add $\vec{x}^r_i$ to the partition list of $\vec{x}^a_i$\;
	}
	
	\Return{$G(V^a, E^a)$, $\mathcal{L}$}\;
	\caption{Naive PAG Construction}
	\label{algo:naive_pag_build}
\end{algorithm}

\subsection{Dynamic Representation Selection}

\begin{figure}[ht!]
	\centering\vspace{-2ex}
	\subfigure[distance exceeds radius]{\includegraphics[width=0.49\linewidth]{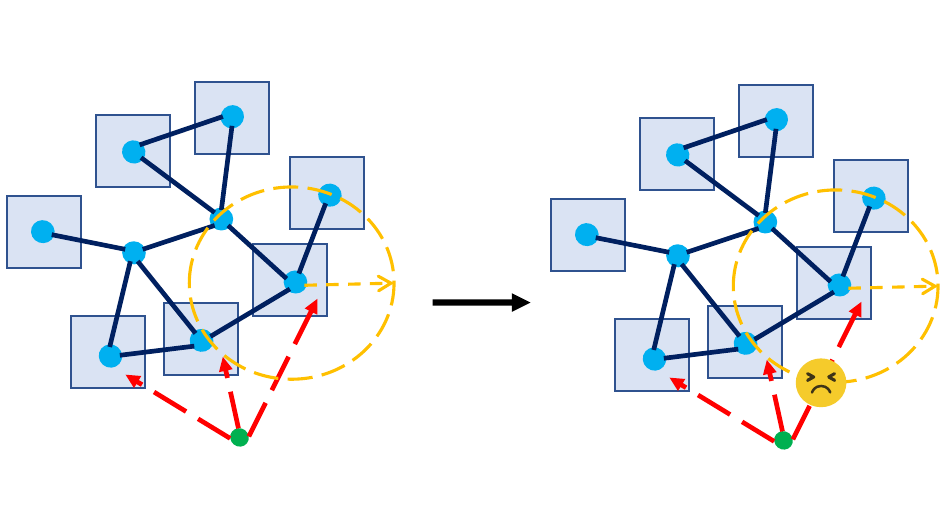}\label{fig:constraint_a}} \hspace{0.2cm}
	\subfigure[promotion]{\includegraphics[width=0.49\linewidth]{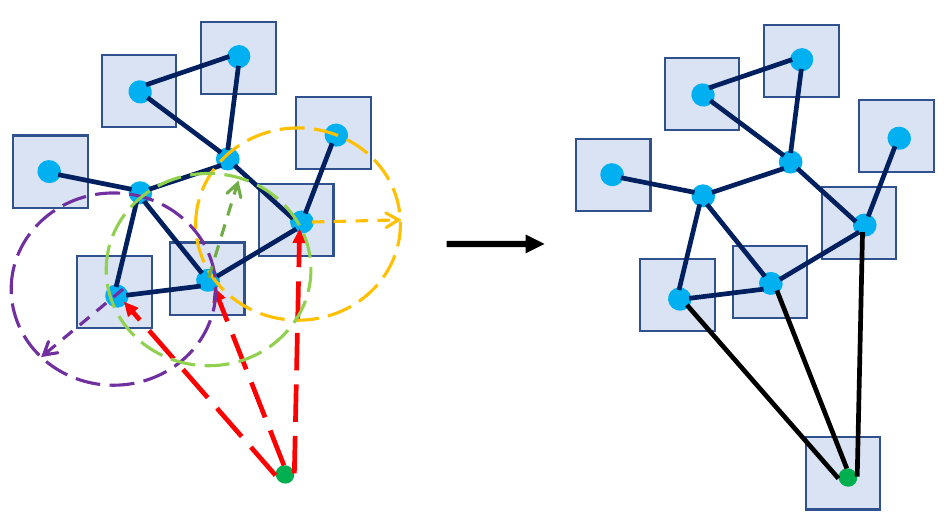}\label{fig:constraint_b}}
	\caption{Illustration of DRS}
	\label{fig:constraint}
\end{figure}

Theoretically, if the ANNS interface provided by the PG $G$ could achieve perfect nearest neighbor retrieval,
the naive PAG could be regarded as a PG constructed on the centroids generated by K-Means algorithm with zero iterations.
However, K-Means typically requires multiple iterations to ensure the quality of clustering 
and relying on zero iterations fail to provide such guarantee \cite{ahmed2020k, DBLP:conf/bigdataconf/LiuHCL0Z18}. 
Furthermore, the straightforward approach of assigning each residual point 
to the partition associated with its nearest neighbor may yield highly imbalanced partitions due to the data skew.
Such imbalance in clustering introduces two primary challenges:
\begin{itemize}[leftmargin=10pt, topsep=1pt]
	\item{} Long Tail Effect: Imbalanced clustering may result in partitions of two extreme sizes.
	Excessively large partitions will significantly increase the query latency of 99.9\%, 
	making it difficult to guarantee the quality of online services. 
	Additionally, we have observed that in certain datasets,
	the largest partition can comprise up to 1\% of the size of $\mathcal{X}$. 
	In the most extreme case, all residual points may be aggregated into a single partition, 
	resulting in the algorithm reducing to exhaustive search.
	\item{} Low Recall Rate: Significant partition imbalance indicates that 
	centroids generated through random sampling may not adequately represent 
	the underlying distribution of data, leading to considerable boundary problem.
	Ideally, the size of each partition should be nearly uniform,
	meanwhile allowing larger partitions exist 
	because the local regions may be a little dense observed from real datasets.
	SPANN also emphasizes the importance of maintaining uniform partitions \cite{DBLP:conf/nips/ChenZWLLLYW21}.
\end{itemize}
To address the challenge of imbalanced partitions, 
we introduce the \emph{Dynamic Representation Selection} (DRS) strategy in this section,
which is summarized in Algorithm \ref{algo:drs}.
(1) DRS explicitly constrains the capacity of each partition to prevent excessively large partitions.
The sample rate $p$ can be conceptually interpreted as the compression rate with
$\frac{1}{p}$ indicating the number of points each partition should ideally preserve. 
For example, if $p = 1\%$, then $\frac{1}{p} = 100$, 
suggesting that each partition should keep approximately 100 points.
To handle data skew,
we allow partitions to moderately exceed their ideal capacity.
Specifically, we set a partition capacity limitation of $\frac{\lambda}{p}, \lambda \ge 1$.
(2) Additionally, the distance limit obtained through distance sampling ensures that 
only residual points within this distance can be aggregated to the corresponding aggregation point,
as illustrated in Figure~\ref{fig:constraint_a}.
The aggregation point, 
along with its residual points, defines a sphere with center 
as described in Definition~\ref{def:pag}.
For adaptive aggregation, 
we calculate the distances between each aggregation point and its neighbors in PG,
then use the $\gamma_1$-th sorted distance as the radius threshold for aggregation.
However, some aggregation points in PG lack sufficient neighbors, 
making the $\gamma_1$ position distance meaningless.
Therefore, we sort radii of all aggregation points 
and use the $\gamma_2$ position radius as the maximum threshold
for every aggregation point.
(3) Finally, residual points that cannot be aggregated into any 
existing aggregation points will be promoted as new aggregation points 
by inserting them into the PG,
as shown in Figure~\ref{fig:constraint_b}.
Through the dynamic promotion, 
we aim to capture the underlying distribution of the dataset.

The DRS method improves aggregation quality through conditional constraints, 
while point promotion provides more options for subsequent point aggregation.

\begin{algorithm}[t]
	\DontPrintSemicolon
	\KwIn{A vector dataset $\mathcal{X}$ of $n$ vectors, sample rate $p$, retrieval neighbor number $k$, 
		capacity extra rate $\lambda$, neighbor distance percentile $\gamma_1$, radius percentile $\gamma_2$}
	\KwOut{A PAG $G(V, E)$ with DRS}
	$c \leftarrow \frac{\lambda}{p}$\;
	
	
	$\mathcal{X}^a \leftarrow$ random sample from $\mathcal{X}$ \;
	\tcp*[l]{keep the fully connected graph as partition}
	$\mathcal{L} \leftarrow$ initialize partition lists of aggregation points \;
	$G_p(V^a, E^a) \leftarrow$ PG on $\mathcal{X}^a$ \;
	
	$\mathcal{D} \leftarrow$ initialize radius map of aggregation points \;
	$\mathcal{L}_\mathcal{D} \leftarrow$ initialize radius list \;
	
	\ForEach{$\vec{x}^a_i \in \mathcal{X}^a$}{
		$L \leftarrow$ sorted distances of $\vec{x}^a_i$ with its neighbors in $G_p$ with length $l$ \;
		$d \leftarrow L[\gamma_1 \times l]$ \;
		$\mathcal{D} \leftarrow \mathcal{D} \cup \{(\vec{x}^a_i, d)\}$ \;
		$\mathcal{L}_\mathcal{D} \leftarrow \mathcal{L}_\mathcal{D} \cup \{d\}$ \;
	}
	
	$d_o \leftarrow$ element at the $\gamma_2$-th percentile position in $\mathcal{L}_\mathcal{D}$ \;
	\ForEach{$(\vec{x}^a_i, d_i) \in \mathcal{D}$} {
		$d_i \leftarrow \min(d_i, d_o)$ \;
	}
	
	\ForEach{$\vec{x}^r_i \in \mathcal{X}^r = \mathcal{X} \backslash \mathcal{X}^a$}{
		$NN(\vec{x}^r_i) \leftarrow$ retrieve $k$ nearest neighbors of $\vec{x}^r_i$ by the greedy search on $G_p(V^a, E^a)$\;
		\ForEach{$\vec{x}^a_j \in NN(\vec{x}^r_i)$}{
			\uIf{$\delta(\vec{x}^r_i, \vec{x}^a_j) \le \mathcal{D}[\vec{x}^a_j]$ and $|\mathcal{L}_j| < c$}{
				add $\vec{x}^r_i$ to partition list $\mathcal{L}_j$ of $\vec{x}^a_j$\;
				\textbf{break}\;
			}
		}
		
		\uIf{$\vec{x}^r_i$ has not been aggregated}{
			insert $\vec{x}^r_i$ into $G$\;
		}
	}
	
	\Return{$G$, $\mathcal{L}$, $\mathcal{D}$}\;
	\caption{PAG Construction with DRS}
	\label{algo:drs}
\end{algorithm}





\subsection{Graph-based Redundancy}

Aggregation and clustering share many similarities and often suffer from the boundary problem \cite{DBLP:conf/nips/ChenZWLLLYW21}. 
To alleviate the boundary problem, one approach explores multiple partitions during the search process 
until the desired performance requirements are met. 
Although this method can slightly improve recall rate, 
it introduces significant latency.
A more straightforward solution is to introduce redundancy by allocating data points to multiple partitions. 
However, determining the selection of redundant partitions is a key challenge, 
as different strategies can lead to various adverse effects:
\begin{itemize}[leftmargin=10pt, topsep=1pt]
	\item{} Dense Redundancy: While we aim to improve index performance, 
	it is essential to minimize repeated access to redundant points.
	When a point's redundant partitions are very close to each other, 
	there is a high probability that the points will be accessed multiple times.
	\item{} Sparse Redundancy: Conversely, unrelated sparse redundancy fails to improve recall, 
	as these redundant points may not be effectively recall during search process.
\end{itemize}
In this paper, we propose a Graph-based Redundancy (GR) strategy that
selects redundant partitions of residual points by using the structural information of the graph, 
thereby avoiding overly dense and sparse redundancy distributions. 
The core idea of the redundancy strategy is to serve the search process. 
Hence, we propose the following two redundancy strategies:
\begin{itemize}[leftmargin=10pt, topsep=1pt]
	\item Nearest Neighbor Redundancy: By utilizing the ANNS interface provided by the PG, 
	we identify the $k$ nearest neighbors of the residual points, 
	and then apply RNG-based rules to select redundant partitions represented by aggregation points.
	\item Routing Path Redundancy: We observe that similar queries often follow the same routing path.
	During the greedy search process of the PG, we track the routing path,
	and then select redundant partitions along the routing path based on RNG rules.
\end{itemize}

We define RNG-based rules as follows \cite{DBLP:journals/pieee/JaromczykT92, DBLP:conf/cvpr/HarwoodD16}.
\begin{definition}
	\label{def:rng}
	\textbf{RNG-base Rules}. 
	Given partition lists $\mathcal{L}$ represent by $\mathcal{X}^a$ and a residual point $\vec{x}^r \in \mathcal{X}^r$,
	aggregation point $\vec{x}^a_1$ occludes aggregation point $\vec{x}^a_2$
	if $\delta(\vec{x}^a_1, \vec{x}^r) < \delta(\vec{x}^a_2, \vec{x}^r)$
	and $\delta(\vec{x}^a_1, \vec{x}^a_2) < \delta(\vec{x}^a_2, \vec{x}^r)$
\end{definition}

\subsection{Concurrent Index Construction}

Although sampling methods can significantly reduce the number of points required for constructing a PG, 
constructing such a graph remains computationally expensive.

In this section, we propose Concurrent Index Construction (CIC) to achieve efficient PG construction.
The details are summarized in Algorithm \ref{algo:cic}.
Following DiskANN methodology, CIC splits the dataset into multiple partitions for independent graph construction
and then merges all individual graphs to a complete graph.
Instead of merging the neighbors of redundant points used by DiskANN, 
CIC let each partition's points identify their nearest neighbors from the close partitions
using ANNS interface provided by the corresponding PG,
and then employ prune strategy to preserve the properties of PG such as KNNG and RNG \cite{DBLP:journals/pieee/JaromczykT92, DBLP:conf/ijcai/HajebiASZ11}.
More importantly, CIC can be easily scaled for distributed construction across multiple machines.

The time complexity associated with native PG construction for $n$ vectors is $O(n\lg n)$ \cite{DBLP:journals/pami/MalkovY20, DBLP:conf/mcpr2/ChavezT10}.
DiskANN replicates points across multiple partition to construct graphs,
achieving a complexity of
\begin{equation}
	O(c \times \frac{\theta n}{c} \lg{\frac{\theta n}{c}}) =  O(\theta n \lg{\frac{\theta n}{c}})
\end{equation}
where $c$ represents the number of partitions
and $\theta$ is the number of replicas.

In contrast, CIC can construct $c$ PGs concurrently, each comprising $\frac{n}{c}$ vectors,
with a time complexity of $O(\frac{n}{c} \lg \frac{n}{c})$.
Additionally, ANNS on a PG which contains $\frac{n}{c}$ vectors will take $O(\lg \frac{n}{c})$.
Consequently, processing all vectors within a partition will require $O((c - 1)\frac{n}{c} \lg \frac{n}{c})$.
Since all partitions can execute ANNS in parallel, the overall time complexity of CIC can be expressed as

\begin{equation}
	O(c \times \frac{n}{c} \lg \frac{n}{c} + c \times (c - 1)\frac{n}{c} \lg \frac{n}{c}) = O(cn \lg \frac{n}{c})
\end{equation}

\begin{algorithm}[t]
	\DontPrintSemicolon
	\KwIn{A vector dataset $\mathcal{X}$ of $n$ vectors, the number of neighbor $k$}
	\KwOut{A PG $G(V, E)$}
	
	$\{ \mathcal{X}^0, \mathcal{X}^1, \ldots, \mathcal{X}^c \} \leftarrow$ split $\mathcal{X}$ into $c$ partitions \;
	
	\tcp*[l]{on parallel}
	\ForEach{$i = 0, \ldots, c$}{
		$G_i \leftarrow$ construct a PG on $\mathcal{X}^i$ 
	}
	
	\tcp*[l]{on parallel}
	\ForEach{$i = 0, \ldots, c$}{
		\ForEach{$\vec{x}^i_s \in \mathcal{X}^i$}{
			$\vec{x}^i_c \leftarrow$ centroid of $\mathcal{X}^i$\; 
			\ForEach{$j = 0, \ldots, c$}{
				$\vec{x}^j_c \leftarrow$ centroid of $\mathcal{X}^j$\; 
				\uIf{$j \neq i$ and $\delta(\vec{x}^i_s, \vec{x}^j_c) \le \eta \delta(\vec{x}^i_s, \vec{x}^i_c)$}{
					$NN(\vec{x}^i_s)_j \leftarrow$ $k$ neighbors of $\vec{x}^i_s$ by ANNS on $G_j$
				}
				\Else{
					$NN(\vec{x}^i_s)_j \leftarrow$ $k$ neighbors of $\vec{x}^i_s$ on $G_j$
				}     
			}
			$NN(\vec{x}^i_s) \leftarrow$ prune neighbors of $\{ NN(\vec{x}^i_s)_0, NN(\vec{x}^i_s)_1, \ldots, NN(\vec{x}^i_s)_c \}$
		}
	}
	
	\Return{$G$ represented by $NN$}\;
	\caption{Concurrent Index Construction}
	\label{algo:cic}
\end{algorithm}

Nevertheless, the cost of ANNS on all partitions is considerable. 
We further enhance CIC efficiency by limiting the number of partitions that each vector searches. 
Ideally, an effective clustering algorithm should constrain proximate points within the same partition
and points residing in adjacent partitions are likely to exhibit greater similarity. 
Therefore, during the graph merge stage, each point needs only to consider its neighboring partitions.
Our experimental results indicate that, 
generally, searching partitions within a distance threshold is 
sufficient to achieve comparable effectiveness to searching all partitions.
Formally, this can be expressed as $\delta(\vec{x}, \vec{x_{c'}}) \le \eta \delta(\vec{x}, \vec{x_c})$,
where $\vec{x_c}$ denotes the centroid of the partition containing $\vec{x}$, 
$\vec{x_{c'}}$ represents the partition under consideration, 
and $\eta$ is the scaling factor.

\section{Search On Index}
\label{sec:search}

\subsection{Query on Index}
During the search process, 
DSANN follows the inverted index methodology \cite{DBLP:conf/cvpr/BabenkoL12}. 
Initially, it traverses the PG stored in memory to identify the most suitable partitions. 
Then, a fine-grained full-scan within the partitions is performed to retrieve the final results.

The number of partitions identified during the graph traversal significantly 
influences both the recall rate and throughput. 
Employing a strategy that probes a fixed number of partitions may lead to 
over-exploration of certain points while under-exploring others \cite{DBLP:conf/osdi/ZhangXCSXCCH00Y23}. 
In this paper, we propose an Adaptive Partition Probe (APP) method, 
which dynamically explores partitions based on the structural information of 
the graph to achieve early stopping.

During the traversal of the PG, 
DSANN maintains the distance information of each visited point, creating a timeline.
The traversal terminates when the distance between the query and current point exceeds $d + r_i + r_j$,
where $d$ represents the minimal distance to the centroid, $r_i$ denotes the aggregation radius of the nearest partition,
and $r_j$ is the aggregation radius of the current partition.

\begin{figure}[ht!]
	\centering
	\includegraphics[width=0.98\linewidth]{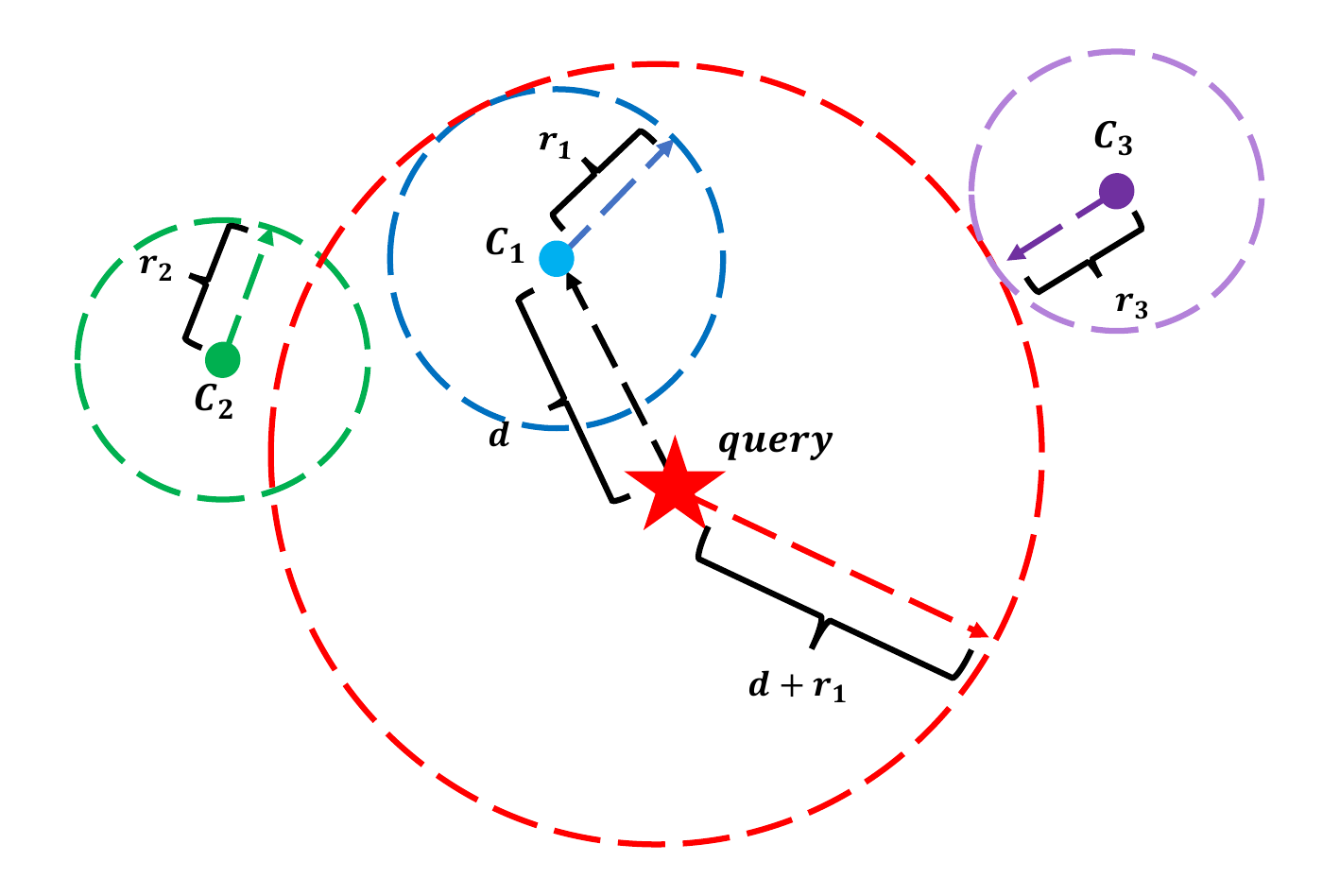}
	\caption{Example of early stop}
	\label{fig:sphere}
\end{figure}

As illustrated in Figure~\ref{fig:sphere}, if the minimal distance between the query and aggregation point is $d$,
the maximum distance within the partition $\mathcal{C}_1$ is $d + r_1$ because the aggregation partition is actually a sphere.
Consider a sphere around the query with radius $d + r_1$,
if there exists an overlap between the partition and the sphere, 
we should explore the partition.
For partition $\mathcal{C}_2$, where the distance between the query and its centroid is less than $d + r_1 + r_2$,
we will explore $\mathcal{C}_2$.
Conversely, we will not explore $\mathcal{C}_3$.

Furthermore, the number of partitions explored is closely related to the $k$, which signifies the number of vectors recalled.
We use $\rho > 1$ as the scale factor for different scenarios.

\subsection{Asynchronous Query}

The inverted index method consists of two stages: a coarse-grained search followed by a fine-grained search.
In the coarse-grained search stage, 
all residual points must be loaded into memory before entering the fine-grained stage. 
Therefore, the coarse-grained stage will stall the latter stage,
particularly in distribution and huge vector retrieval scenarios, 
where it may result in significant I/O and computational imbalance
because distribution storage introduces extra network I/O latency
and huge vector retrieval means that we need to fetch a lot of residual points. 

In this paper, we introduce a search algorithm with asynchronous I/O and computation, 
as summarized in Algorithm \ref{algo:search}.
Leveraging the structural information of PG,
DSANN can achieve asynchronous graph traversal in memory 
while concurrently retrieving the qualified partitions into memory for full-scan. 
By controlling the strictness of fetching criteria, 
we can effectively balance I/O and computation, improving the throughput of the algorithm.
For example, we can fetch the residual points as soon as 
the aggregation point with minimal distance is probed.

Although caching with local memory and disk can effectively reduce I/O overhead in distributed storage, 
the search pattern of DSANN, similar to SPANN, introduces unpredictability in partition access,
complicating the identification of frequently accessed partitions prior to query execution.
Consequently, the effectiveness of caching is significantly constrained.
Nevertheless, optimizing caching strategies remains a promising direction for future research.

\begin{algorithm}[t]
	\DontPrintSemicolon
	\KwIn{A PAG $G(E, V)$ with partition list $\mathcal{L}$, radius map $\mathcal{D}$,
		start node $s$, query $q$,
		distance scale factor $\rho$
		and result size $k$}
	\KwOut{Result set $\mathcal{R}$ containing 
		$k$ approximate nearest neighbors of $q$}
	
	initialize timeline of visited aggregation points $\mathcal{T} \leftarrow$ $\{s \}$ \;
	initialize visited aggregation points set $\mathcal{V} \leftarrow$ $\emptyset$ \;
	initialize partitions has been loaded into memory $\mathcal{C} \leftarrow \emptyset$ \;
	
	\While{$\mathcal{T} \backslash \mathcal{V} \neq \emptyset$}{
		$p^* \leftarrow$ $\argmin_{p \in \mathcal{T} \backslash \mathcal{V}}{\delta(p, q)}$ \;
		$\mathcal{T} \leftarrow \mathcal{T} \cup NN(p^*)$ \;
		$\mathcal{V} \leftarrow \mathcal{V} \cup p^*$ \;
		
		\If{$\mathcal{T}$ has reached local optimum $v^*$}{
			\If{$\delta(q, p^*) \ge \rho (\delta(v^*, p^*) + \mathcal{D}[v^*] + \mathcal{D}[p^*])$}{
				\textbf{break}
			}
			
			\tcp*[l]{asynchronous I/O}
			$c^* \leftarrow \argmin_{c \in \mathcal{T} \backslash \mathcal{C}}{\delta(c, q)}$ \;
			$\mathcal{C} \leftarrow \mathcal{C} \cup c^*$ \;
			$\mathcal{L}(c^*) \leftarrow$ load residual points of $c^*$ by I/O \;
			$\mathcal{R} \leftarrow \mathcal{R} \cup \mathcal{L}(c^*)$ \;
		}
	}
	
	\Return{closest $k$ points from $\mathcal{R}$}\;
	\caption{Asynchronous Search On Index}
	\label{algo:search}
\end{algorithm}

\section{Experimental Study}
\label{sec:experimental}

In this section, we present detailed analysis of extensive experiments on public datasets.
The evaluation seeks to answer the following questions:
\begin{itemize}[leftmargin=*]
	\item How do DSANN and current state-of-the-art ANNS algorithms perform in terms of accuracy, performance and resource usage?
	\item How do different strategies contribute to DSANN?
\end{itemize}
\subsection{Experimental Setup}
\noindent\textbf{Datasets}. We use 5 public datasets to comprehensively evaluate the performance. 
The statistics of datasets is summarized in Table~\ref{tab:dataset}.
\begin{itemize}[leftmargin=*]
	\item SIFT is widely used for evaluating the performance of ANNS algorithms, which contains 1M SIFT vectors with 128 dimensions.
	\item GIST is a classical image dataset which contains 1M vectors with 960 dimensions.
	\item Glove is a widely used dataset for natural language processing tasks, consisting of pre-trained word embeddings. 
	It contains 1.2M vectors with 100 dimensions, representing words and their semantic meanings in vector space.
	\item BigANN is a popular dataset for evaluating the performance of ANNS algorithms that can scale to very large datasets,
	which contains 1B vectors with 128 dimensions.
	\item DEEP1B includes the learned features from GoogLeNet model which contains 1B vectors with 96 dimensions.
\end{itemize}

\begin{table}[ht!]
	\setlength{\abovecaptionskip}{0.1cm}
	\caption{Dataset Statistics}
	\label{tab:dataset}
	\centering
	\begin{tabular}{l|c|c|l}
		\hline
		\textbf{Dataset}						& \textbf{Dimensions}					& \textbf{Base} (M:$10^6$, B:$10^9$)					& \textbf{Query}      \\ \hline\hline
		{SIFT}                      & 128                      		& 1M																					& 10,000              \\ \hline
		{GIST}     									& 960                         & 1M                  												& 1,000								\\ \hline
		{Glove}                 		& 100                       	& 1.2M                 												& 1,000								\\ \hline
		{BigANN}                 		& 128		                      & 1B                      										& 10,000						  \\ \hline
		{Deep1B}                    & 96                          & 1B 																					& 10,000              \\ \hline
	\end{tabular}
\end{table}

\subsection{Approaches and Measurements}
Compared Algorithms. We mainly focus on the hybrid storage ANNS algorithms which support large scale datasets including:
(1) SPANN follows the inverted index methodology which stores the centroid points in the memory while putting the large partition list in the disk.
(2) DiskANN stores the PQ compressed vectors in the memory while keeping the navigating spread-out graph along with the full-precision vectors on the disk.
Additionally, we have implemented the Pangu DFS version of the above algorithms for comparison.
Pangu DFS serves as the foundational infrastructure for the object storage service in Alibaba Cloud.

\noindent\textbf{Evaluation Metrics.} The efficiency and effectiveness of index are evaluated by Queries Per Second (QPS) and \emph{recall@$k$}, which are widely used in ANNS.
In particular, QPS is the ratio of the number of queries to the query time, and \emph{recall@$k$} is defined by Equation~\ref{eq:recall}.
What's more, We evaluate the construction efficiency of the above index by build time. 

\noindent\textbf{Implementation Details.} Memory and disk scenario experiments were conducted on the same Linux servers with Intel Xeon Gold Processor at 2.7GHz and 1024G memory.
Distributed storage scenario experiments were conducted on Pangu DFS.
The code of all comparison methods can be publicly accessed in their own GitHub repositories.

\noindent\textbf{Parameters.} We use the full memory to build index
and record both memory usage and build time.
In the in-memory scenario, we compare the performance of PAG and DiskANN 
while ensuring that the index size remains identical (SPANN does not have an in-memory version). 
For PAG, we set the sample rate to 20\% and the degree of the graph to 16.
In the disk-based and DFS scenarios, we constrain the memory size to 32G 
and compare the performance of the indices using the parameter configurations 
recommended in the respective comparative algorithm papers.

\subsection{Efficiency and Effectiveness Evaluation}

\begin{figure*}[ht!]\centering\vspace{-2ex}
	\begin{tabularx}{\textwidth}{XXXX}
		\begin{minipage}[c]{.5\textwidth}
			\subfigure[][{\scriptsize QPS v.s. Recall@10}]{
				\scalebox{0.35}[0.35]{\includegraphics{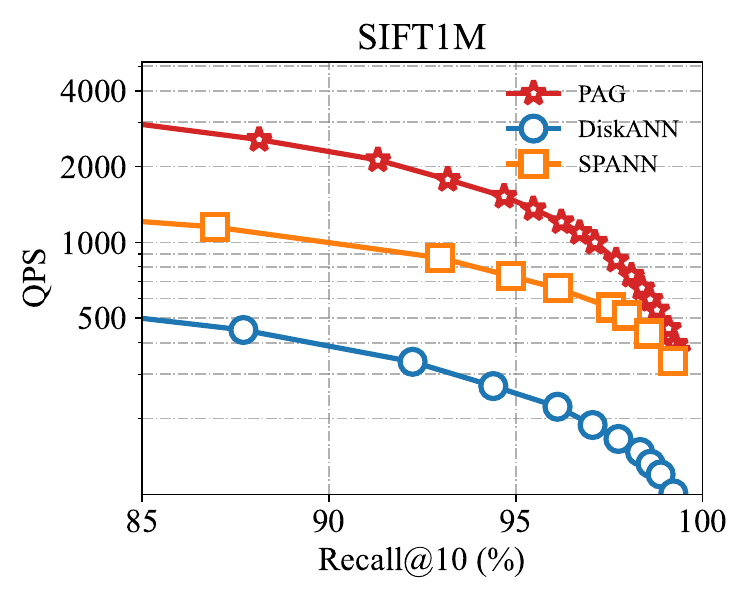}}
				\label{subfig:exp_disk_sift_recall10_qps}}\hspace{-0.5ex}
			
			\subfigure[][{\scriptsize QPS v.s. Recall@100}]{
				\scalebox{0.35}[0.35]{\includegraphics{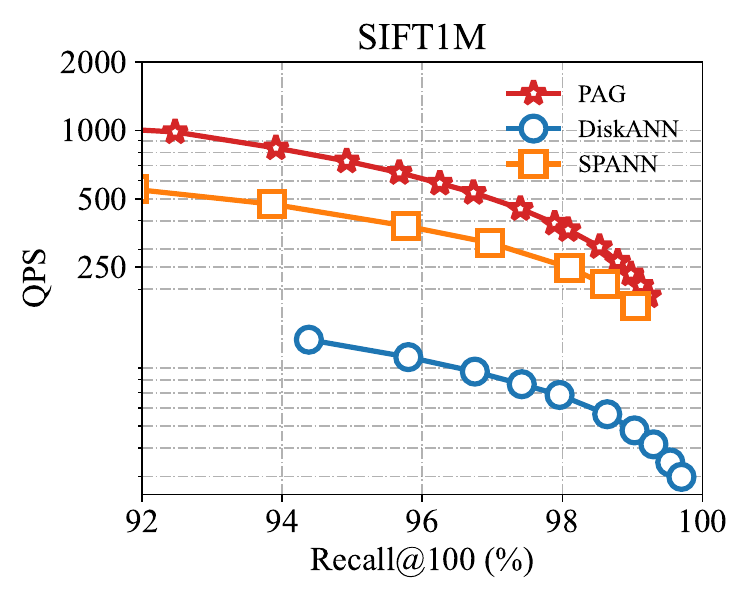}}
				\label{subfig:exp_disk_sift_recall100_qps}}\hspace{-0.5ex}
		\end{minipage} &
		\begin{minipage}[c]{.5\textwidth}
			\subfigure[][{\scriptsize QPS v.s. Recall@10}]{
				\scalebox{0.35}[0.35]{\includegraphics{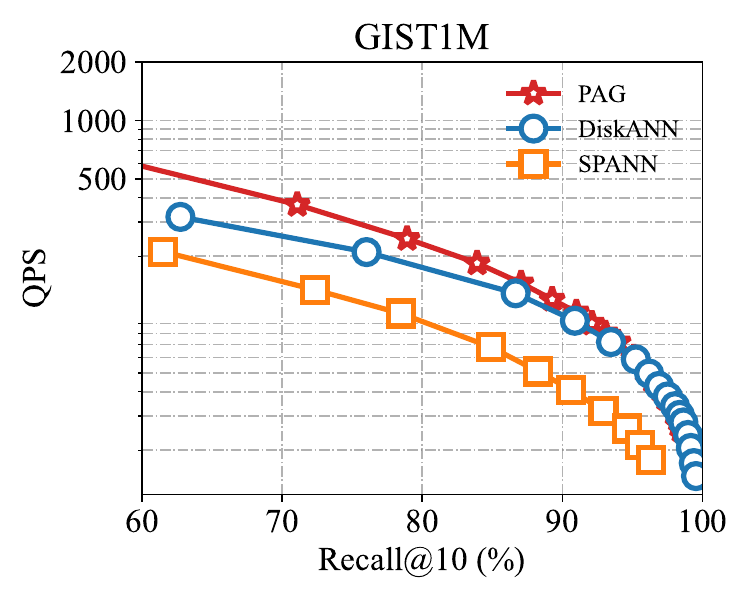}}
				\label{subfig:exp_disk_gist_recall10_qps}}\hspace{-0.5ex}
			
			\subfigure[][{\scriptsize QPS v.s. Recall@100}]{
				\scalebox{0.35}[0.35]{\includegraphics{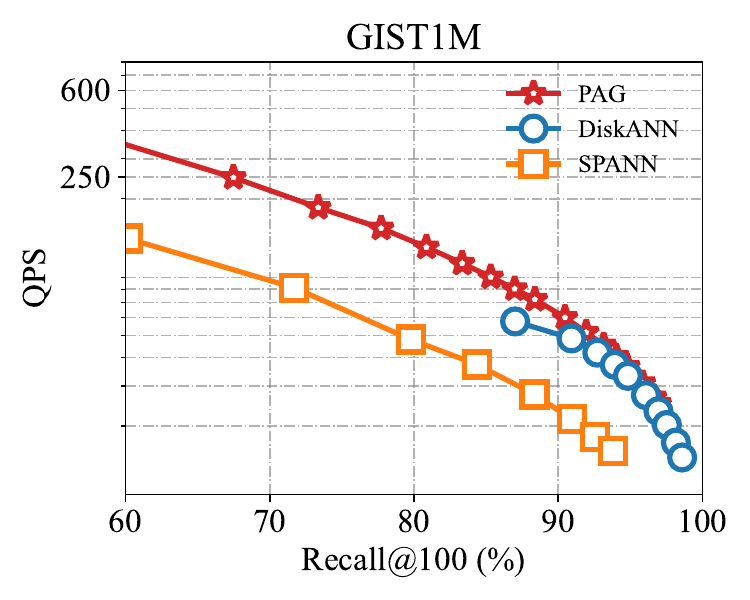}}
				\label{subfig:exp_disk_gist_recall100_qps}}\hspace{-0.5ex}
		\end{minipage} &
		\begin{minipage}[c]{.5\textwidth}
			\subfigure[][{\scriptsize QPS v.s. Recall@10}]{
				\scalebox{0.35}[0.35]{\includegraphics{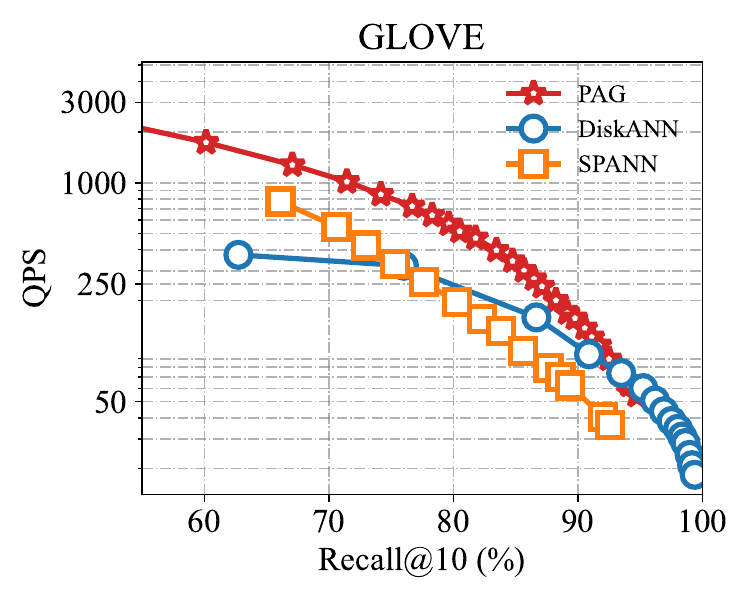}}
				\label{subfig:exp_disk_glove_recall10_qps}}\hspace{-0.5ex}
			
			\subfigure[][{\scriptsize QPS v.s. Recall@100}]{
				\scalebox{0.35}[0.35]{\includegraphics{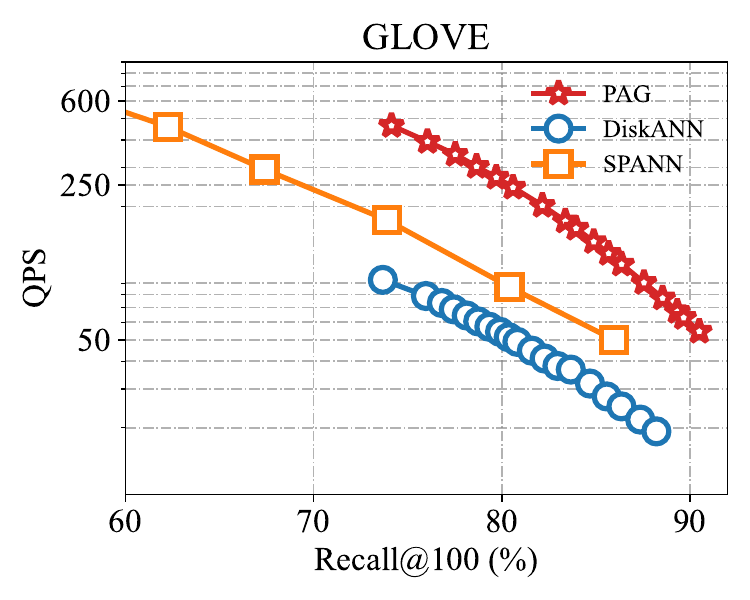}}
				\label{subfig:exp_disk_glove_recall100_qps}}\hspace{-0.5ex}
		\end{minipage} &
		\begin{minipage}[c]{.5\textwidth}
			\subfigure[][{\scriptsize QPS v.s. Recall@10}]{
				\scalebox{0.35}[0.35]{\includegraphics{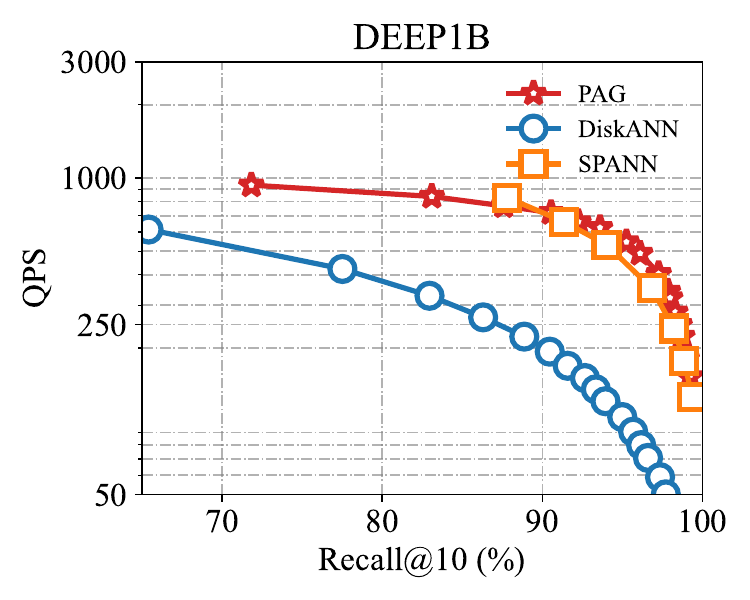}}
				\label{subfig:exp_disk_deep_recall10_qps}}\hspace{-0.5ex}
			
			\subfigure[][{\scriptsize QPS v.s. Recall@100}]{
				\scalebox{0.35}[0.35]{\includegraphics{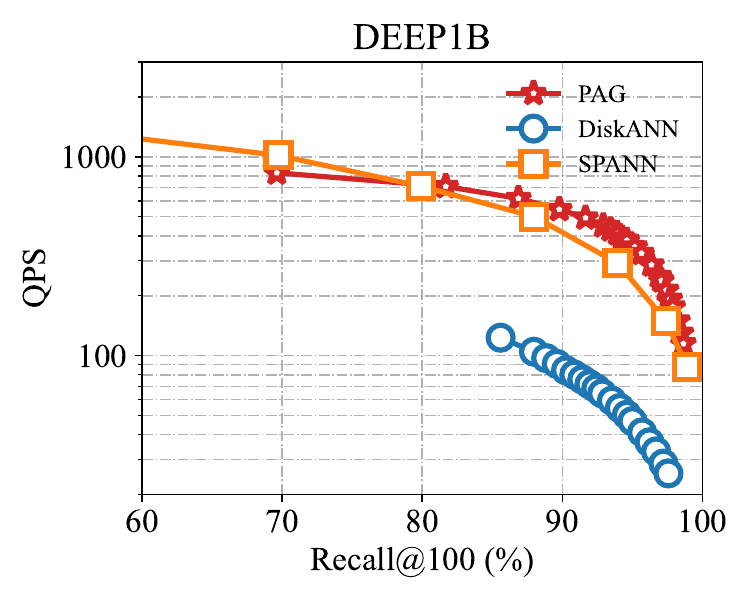}}
				\label{subfig:exp_disk__recall100_qps}}\hspace{-0.5ex}
		\end{minipage}
	\end{tabularx}\vspace{-1ex}
	\caption{Disk-Memory Hybrid Time-Accuracy Trade-off (Varying Algorithms and $k$).}
	\label{fig:exp_disk_recall_qps}
\end{figure*}

\noindent\textbf{Disk-memory hybrid scenario.}
Figure~\ref{fig:exp_disk_recall_qps} presents QPS vs. Recall@$k$ results
for disk-memory hybrid ANNS under identical memory constraints,
comparing PAG (ours), DiskANN, and SPANN.
Each column in Figure~\ref{fig:exp_disk_recall_qps} corresponds to a different dataset,
and each row represents a different $k$ range from 10 to 100.
Our evaluations show that PAG outperforms state-of-the-art disk-memory hybrid
solutions on easy datasets.
For instance, at 95\% Recall@10 in SIFT, 
PAG achieves a QPS of 1359, which is 405\% improvement (or 5.05$\times$ faster) over that 
of DiskANN with a QPS of 269 and
85\% improvement (or 1.85$\times$ faster) over that of SPANN with a QPS of 734.
The performance results for Gist and Glove datasets further demonstrate that 
PAG outperforms SPANN in both recall@10 and recall@100.
This is because SPANN strictly demands the even partition sizes,
which may break the data distribution,
while PAG's DRS can effectively handle such imbalanced datasets.

\begin{figure*}[ht!]\centering\vspace{-2ex}
	\begin{tabularx}{\textwidth}{XXX}
		\begin{minipage}[c]{.5\textwidth}
			\subfigure[][{\scriptsize QPS v.s. Recall@10}]{
				\scalebox{0.4}[0.35]{\includegraphics{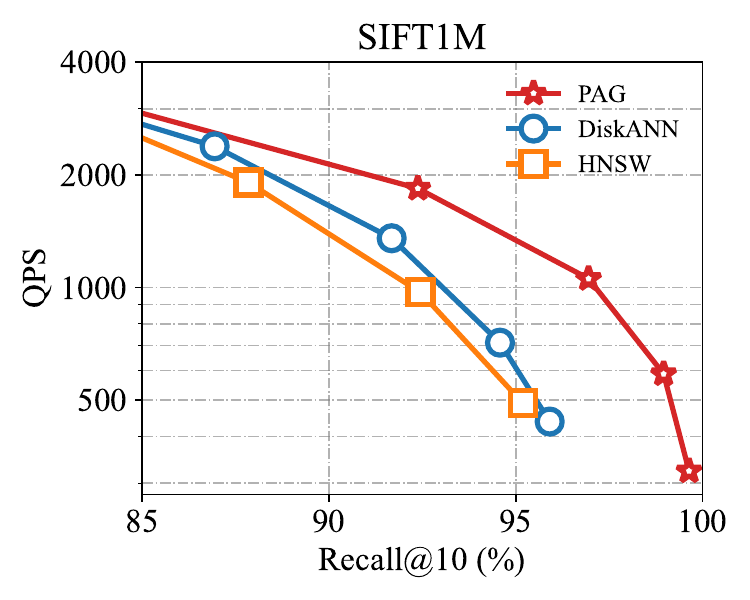}}
				\label{subfig:exp_mem_sift_recall10_qps}}\hspace{-1ex}
			
			\subfigure[][{\scriptsize QPS v.s. Recall@100}]{
				\scalebox{0.4}[0.35]{\includegraphics{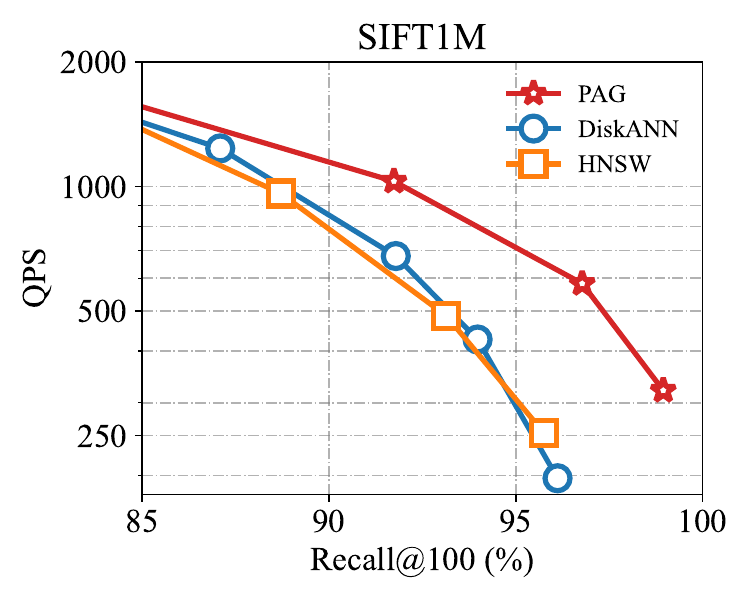}}
				\label{subfig:exp_mem_sift_recall100_qps}}\hspace{-1ex}
			
			\subfigure[][{\scriptsize QPS v.s. Recall@1000}]{
				\scalebox{0.4}[0.35]{\includegraphics{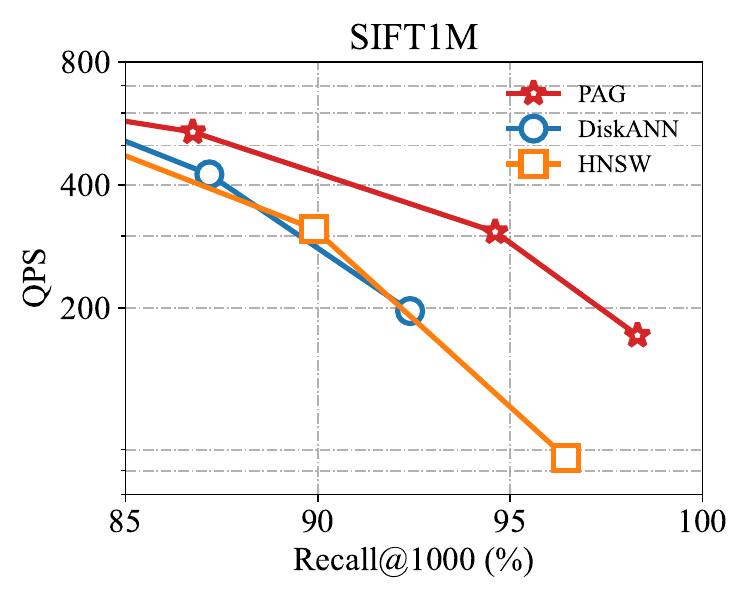}}
				\label{subfig:exp_mem_sift_recall1000_qps}}\hspace{-1ex}
			
			\subfigure[][{\scriptsize QPS v.s. Recall@3000}]{
				\scalebox{0.4}[0.35]{\includegraphics{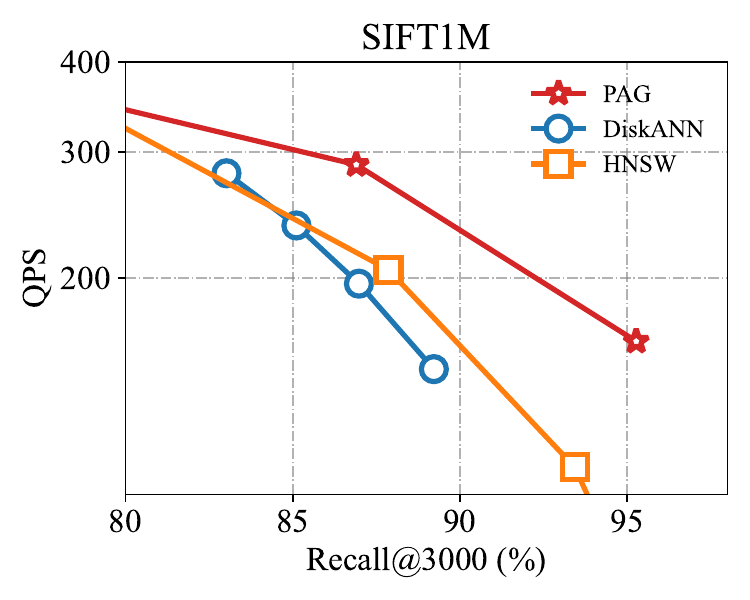}}
				\label{subfig:exp_mem_sift_recall3000_qps}}\hspace{-1ex}
		\end{minipage} &
		\begin{minipage}[c]{.5\textwidth}
			\subfigure[][{\scriptsize QPS v.s. Recall@10}]{
				\scalebox{0.4}[0.35]{\includegraphics{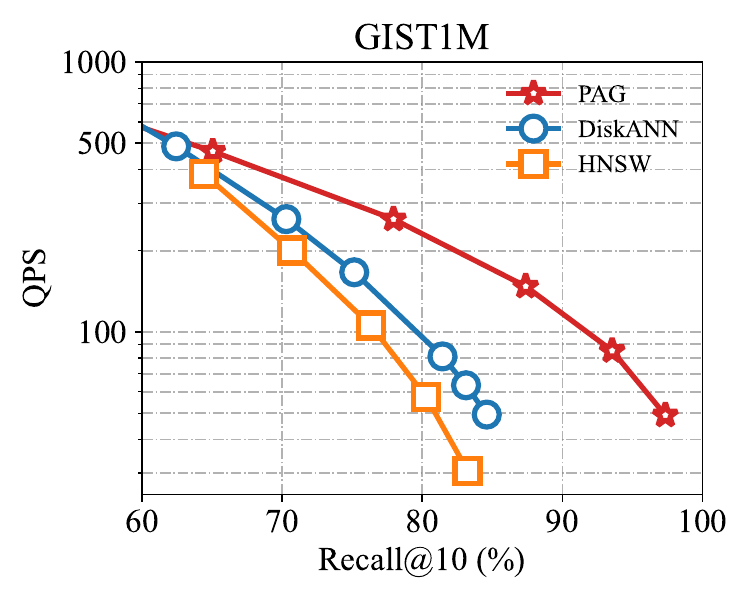}}
				\label{subfig:exp_mem_gist_recall10_qps}}\hspace{-1ex}
			
			\subfigure[][{\scriptsize QPS v.s. Recall@100}]{
				\scalebox{0.4}[0.35]{\includegraphics{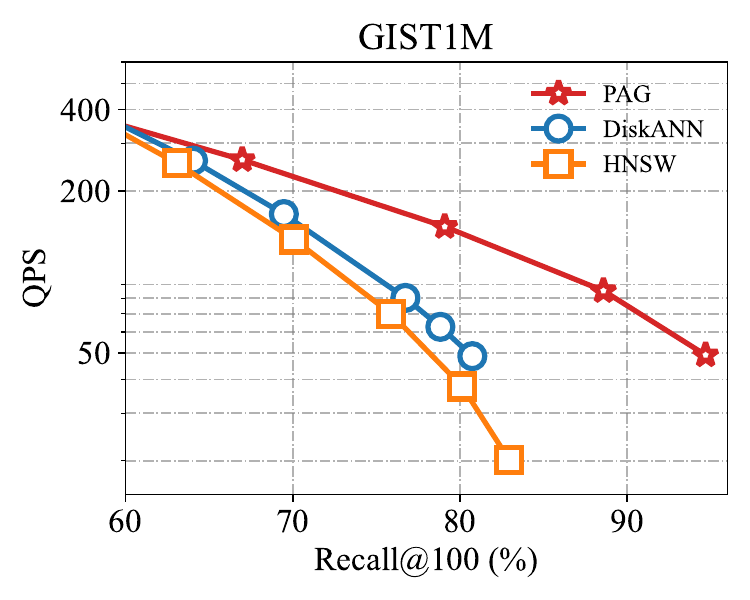}}
				\label{subfig:exp_mem_gist_recall100_qps}}\hspace{-1ex}
			
			\subfigure[][{\scriptsize QPS v.s. Recall@1000}]{
				\scalebox{0.4}[0.35]{\includegraphics{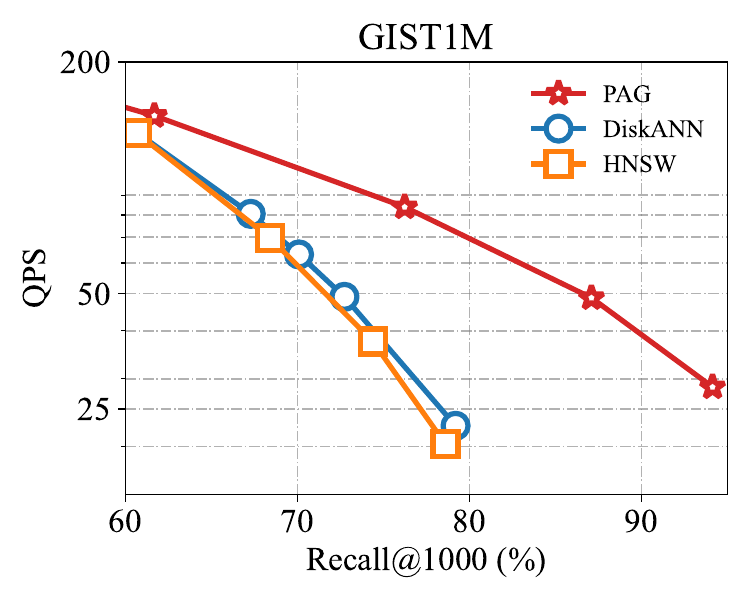}}
				\label{subfig:exp_mem_gist_recall1000_qps}}\hspace{-1ex}
			
			\subfigure[][{\scriptsize QPS v.s. Recall@3000}]{
				\scalebox{0.4}[0.35]{\includegraphics{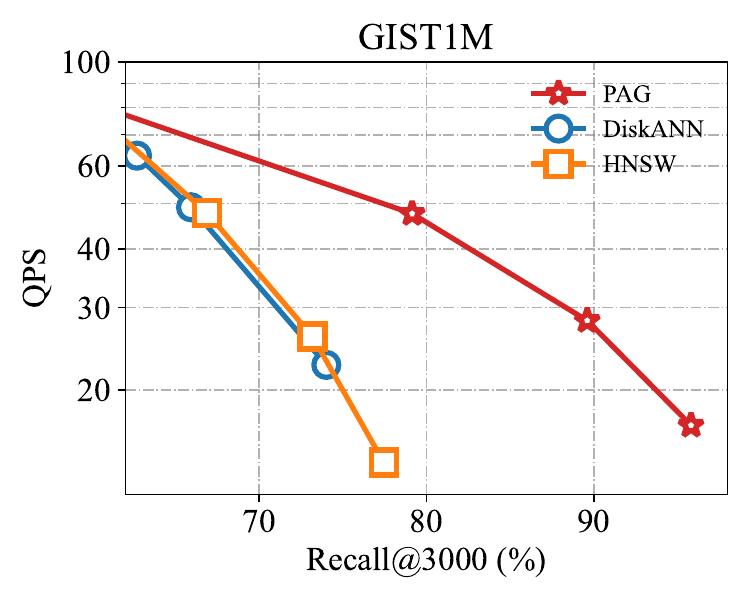}}
				\label{subfig:exp_mem_gist_recall3000_qps}}\hspace{-1ex}
		\end{minipage} &
		\begin{minipage}[c]{.5\textwidth}
			\subfigure[][{\scriptsize QPS v.s. Recall@10}]{
				\scalebox{0.4}[0.35]{\includegraphics{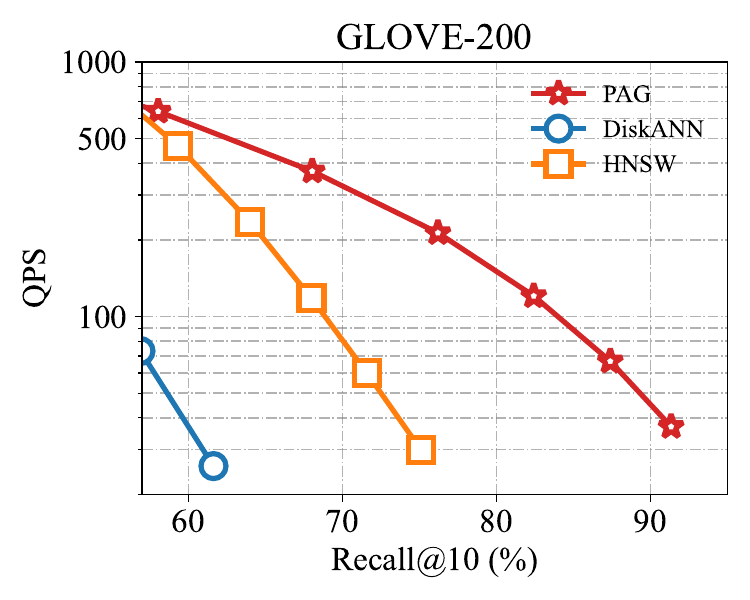}}
				\label{subfig:exp_mem_glove_recall10_qps}}\hspace{-1ex}
			
			\subfigure[][{\scriptsize QPS v.s. Recall@100}]{
				\scalebox{0.4}[0.35]{\includegraphics{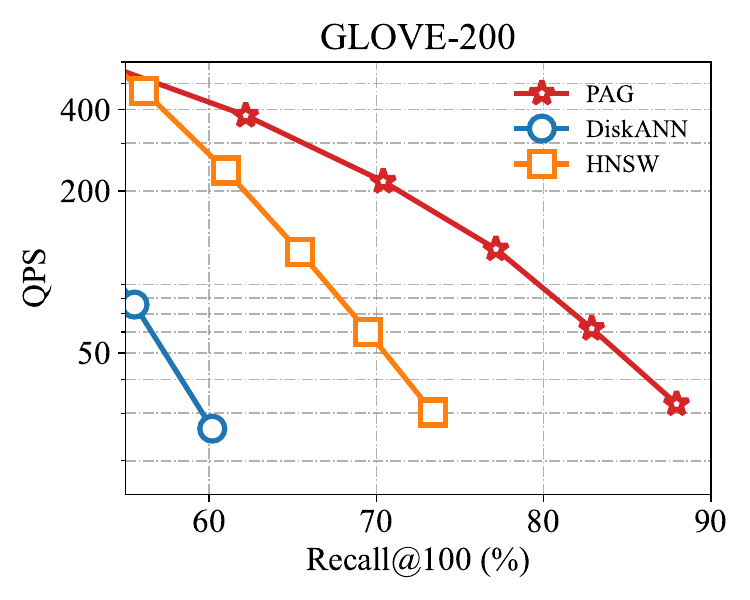}}
				\label{subfig:exp_mem_glove_recall100_qps}}\hspace{-1ex}
			
			\subfigure[][{\scriptsize QPS v.s. Recall@1000}]{
				\scalebox{0.4}[0.35]{\includegraphics{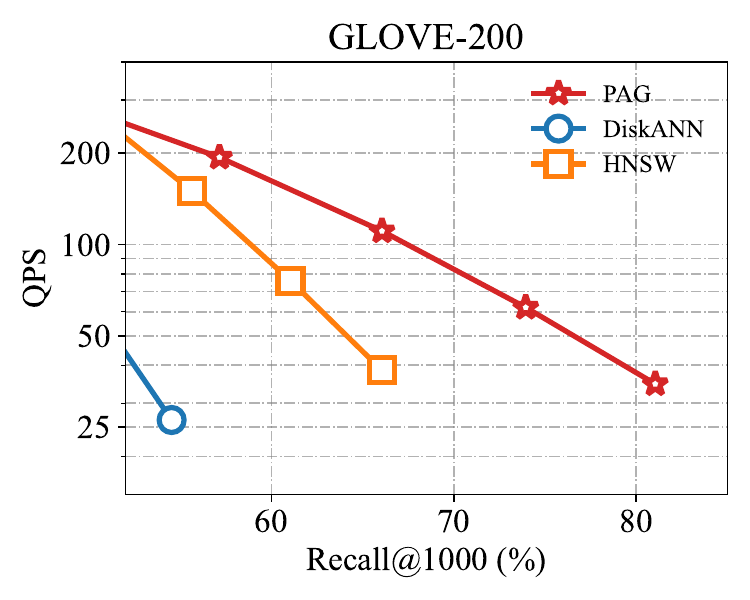}}
				\label{subfig:exp_mem_glove_recall1000_qps}}\hspace{-1ex}
			
			\subfigure[][{\scriptsize QPS v.s. Recall@3000}]{
				\scalebox{0.4}[0.35]{\includegraphics{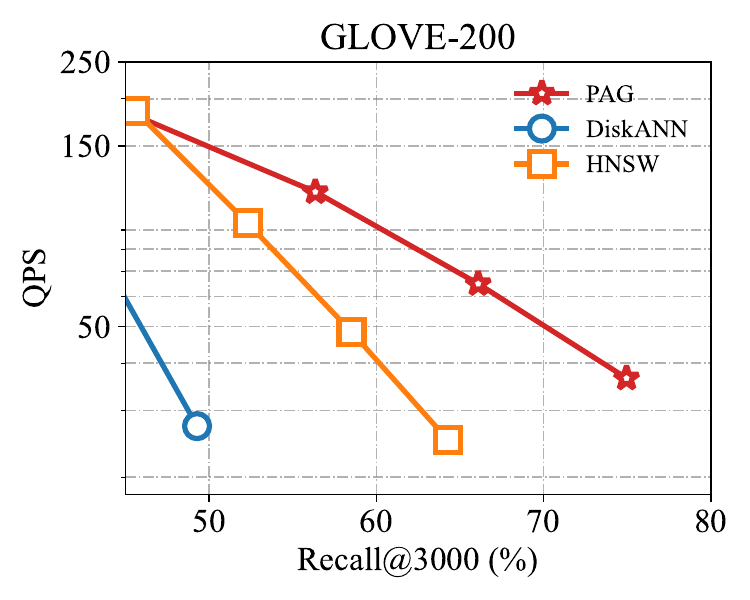}}
				\label{subfig:exp_mem_glove_recall3000_qps}}\hspace{-1ex}
		\end{minipage}
	\end{tabularx}\vspace{-1ex}
	\caption{Memory-only Time-Accuracy Trade-off (Varying Algorithms and $k$).}
	\label{fig:exp_memory_recall_qps}\vspace{-2ex}
\end{figure*}

\noindent\textbf{In memory scenario.} Figure~\ref{fig:exp_memory_recall_qps}
reports the efficiency (QPS) vs. effectiveness (Recall@$k$) results
for memory-only ANNS under the \emph{constraint of the same index size} including PAG (ours), DiskANN and HNSW.
Each column in Figure~\ref{fig:exp_memory_recall_qps} corresponds to a different dataset,
and each row represents a different $k$ range from 10 to 3000.
All experiments were carried out using 1 thread.
Our evaluations show that PAG outperforms DiskANN almost across all $k$ scenarios and datasets.
For example, given the same Recall@10 at 96\% in SIFT, 
PAG achieves a QPS of 1255, which is 186\% improvement (or 2.86$\times$ faster) over that 
of DiskANN with a QPS of 438.
The QPS improvement for Gist are 133\%.
This can be explained by the fact that PAG combines the advantage of 
graph and partition.
Besides, since PAG captures the underlying data distribution,
PAG can well support the imbalanced datasets such as Gist and Glove.

Moreover, we present different Recall@$k$ vs. QPS comparisons
across all datasets in Figure~\ref{fig:exp_memory_recall_qps}.
QPS decreases as $k$ increases,
because more routing steps are needed to retrieve more accurate results.
More hops lead to longer processing time and lower QPS.
PAG retrieves close vectors through partition full-scan with larger fan-out,
which leads to a slower QPS decrease compared to DiskANN.

\begin{figure}[ht!]
	\centering\vspace{-2ex}
	\includegraphics[width=0.8\linewidth]{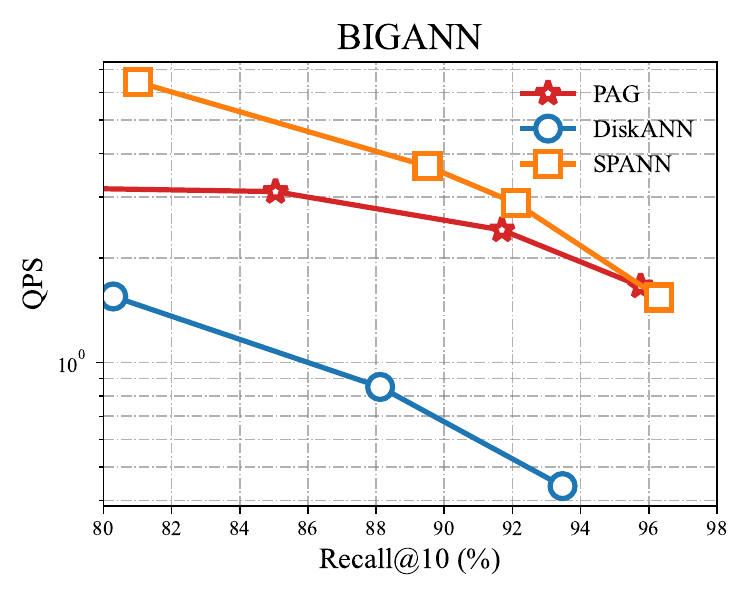}
	\caption{DFS-Memory Hybrid Time-Accuracy Trade-off (Varying Algorithms).}
	\label{fig:exp_dfs_recall_qps}
\end{figure}
\noindent\textbf{DFS-memory hybrid scenario.}
Figure~\ref{fig:exp_dfs_recall_qps} presents QPS vs. Recall@10 results
for DFS-memory hybrid ANNS under identical memory constraints,
comparing PAG (ours) with DiskANN, and SPANN.
Our evaluations demonstrate that PAG significantly outperforms DiskANN,
while achieving comparable performance to SPANN at high recall rates.

\begin{table}[ht!]
	\setlength{\abovecaptionskip}{0.1cm}
	\centering\vspace{-2ex}
	\caption{Build time (seconds)}
	\label{tab:buildcost}
	\renewcommand\arraystretch{1.3}
	\begin{tabular}{c||c|c|c|c}
		\hline
		\textbf{Methods} & \textbf{SIFT} & \textbf{GIST} & \textbf{Glove} & \textbf{Deep}  \\
		\hline
		\hline
		DiskANN            & 32 & 380 & 127 & 632 \\
		\hline 
		SPANN            & 480 & 1052 & 498 & 2080 \\
		\hline 
		PAG            & 20 & 140 & 66 & 305 \\
		\hline 
	\end{tabular}
\end{table}
\noindent\textbf{Build time.}
Table~\ref{tab:buildcost} presents the index build times (in seconds) for PAG (ours), 
DiskANN, and SPANN across different datasets. 
All index constructions are performed using 32 threads. 
The results show that PAG consistently requires less time for index construction compared to both DiskANN and SPANN on all datasets. 
Specifically, for the Deep dataset, PAG requires 305 seconds, while DiskANN and SPANN take 632 and 2080 seconds, respectively. 
PAG is approximately 2 times faster than DiskANN and about 6.8 times faster than SPANN for this dataset.
Similar trends are observed across other datasets, with PAG demonstrating lower build times compared to the other methods.

\subsection{Parameter Analysis}

\begin{figure}[ht!]
	\centering\vspace{-2ex}
	\subfigure[QPS v.s. Recall@10]{\includegraphics[width=0.49\linewidth]{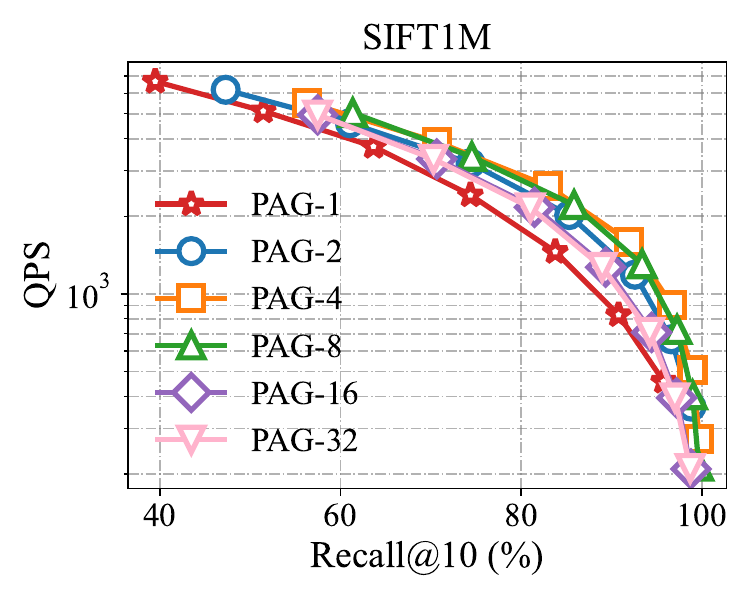}\label{fig:exp_redudant_sift_recall10_qps}} 
	\subfigure[QPS v.s. Recall@10]{\includegraphics[width=0.49\linewidth]{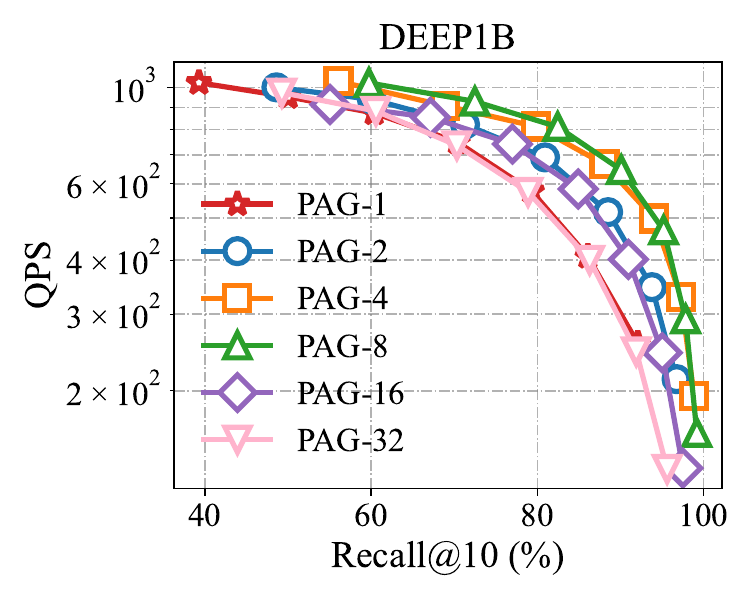}\label{fig:exp_redudant_deep_recall10_qps}}
	\caption{ Effect of redundancy number for Disk-memory hybrid scenario}
	\label{fig:exp_disk_replica}
\end{figure}

\noindent\underline{Effect of redundancy number}
We study the effect of redundancy number on ANNS's performance (QPS vs. Recall@10) 
in Figure~\ref{fig:exp_disk_replica}.
The index performance improvement is most significant 
when the redundancy numbers are set to 4 and 8. 
Lower redundancy numbers are insufficient to effectively improve partition hit rates, 
while excessive redundancy can degrade index performance due to overly dense data.

\begin{figure}[ht!]
	\centering\vspace{-2ex}
	\includegraphics[width=0.98\linewidth]{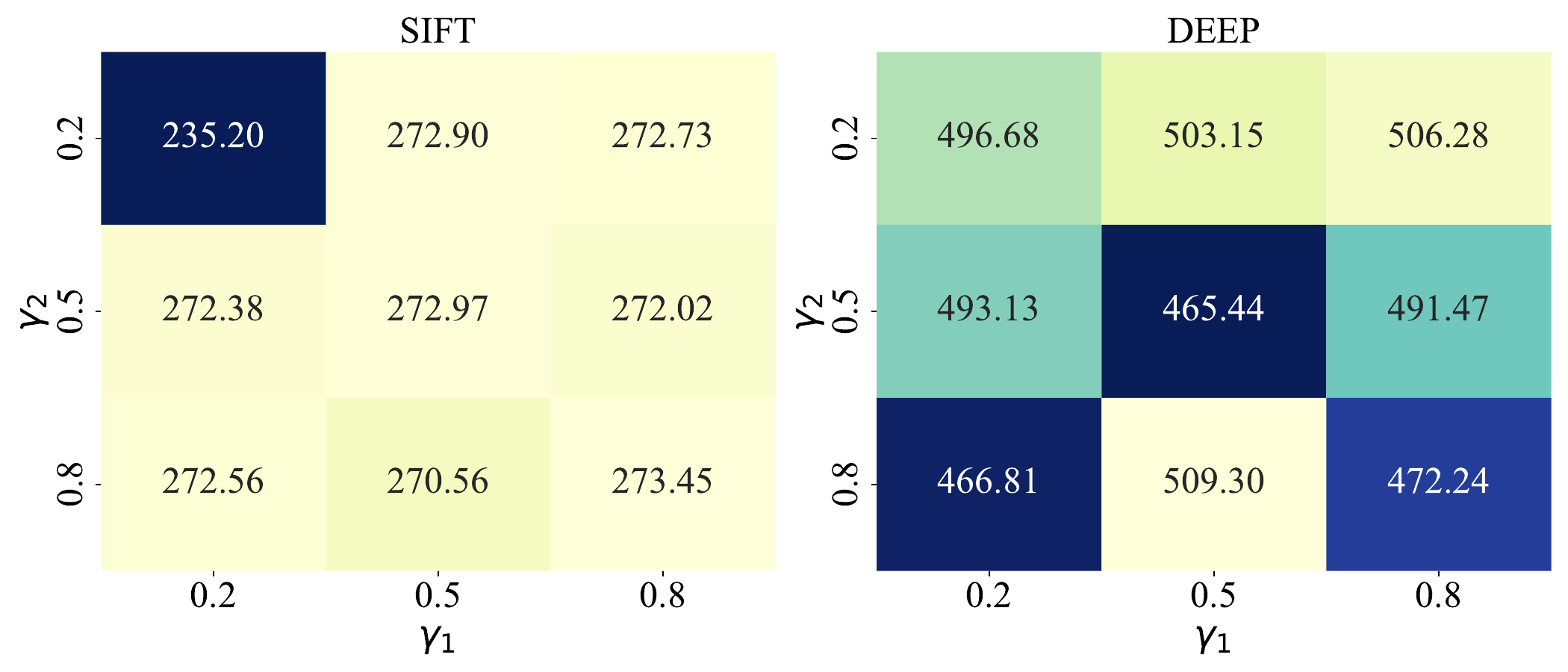}
	\caption{ Effect of radius proportion for Disk-memory hybrid scenario}
	\label{fig:exp_disk_radius}
\end{figure}
\noindent\underline{Effect of radius proportion}
Since the sampling ratios of radius, $\gamma_1$ and $\gamma_2$, 
are critical to the aggregation radius size, Figure~\ref{fig:exp_disk_radius} illustrates 
the impact of $\gamma_1$ and $\gamma_2$ on ANNS performance across different datasets. 
Each value in the grid represents the QPS achieved at a Recall@10 of 99\% and 93\% 
for a specific pair of $\gamma_1$ and $\gamma_2$. 
Overall, the index performance is not significantly affected by these parameters. 
However, larger $\gamma_1$ and smaller $\gamma_2$ yield the most noticeable improvements in index performance. 
This can be explained by the fact that a larger $\gamma_1$ helps aggregate nearby points, 
while a smaller $\gamma_2$ helps avoid the occurrence of extreme values. 
In contrast, smaller $\gamma_1$ values, due to overly strict aggregation conditions, 
fail to effectively group points, resulting in slightly worse index performance compared to other configurations.

\begin{figure}[ht!]
	\centering\vspace{-2ex}
	\includegraphics[width=0.90\linewidth]{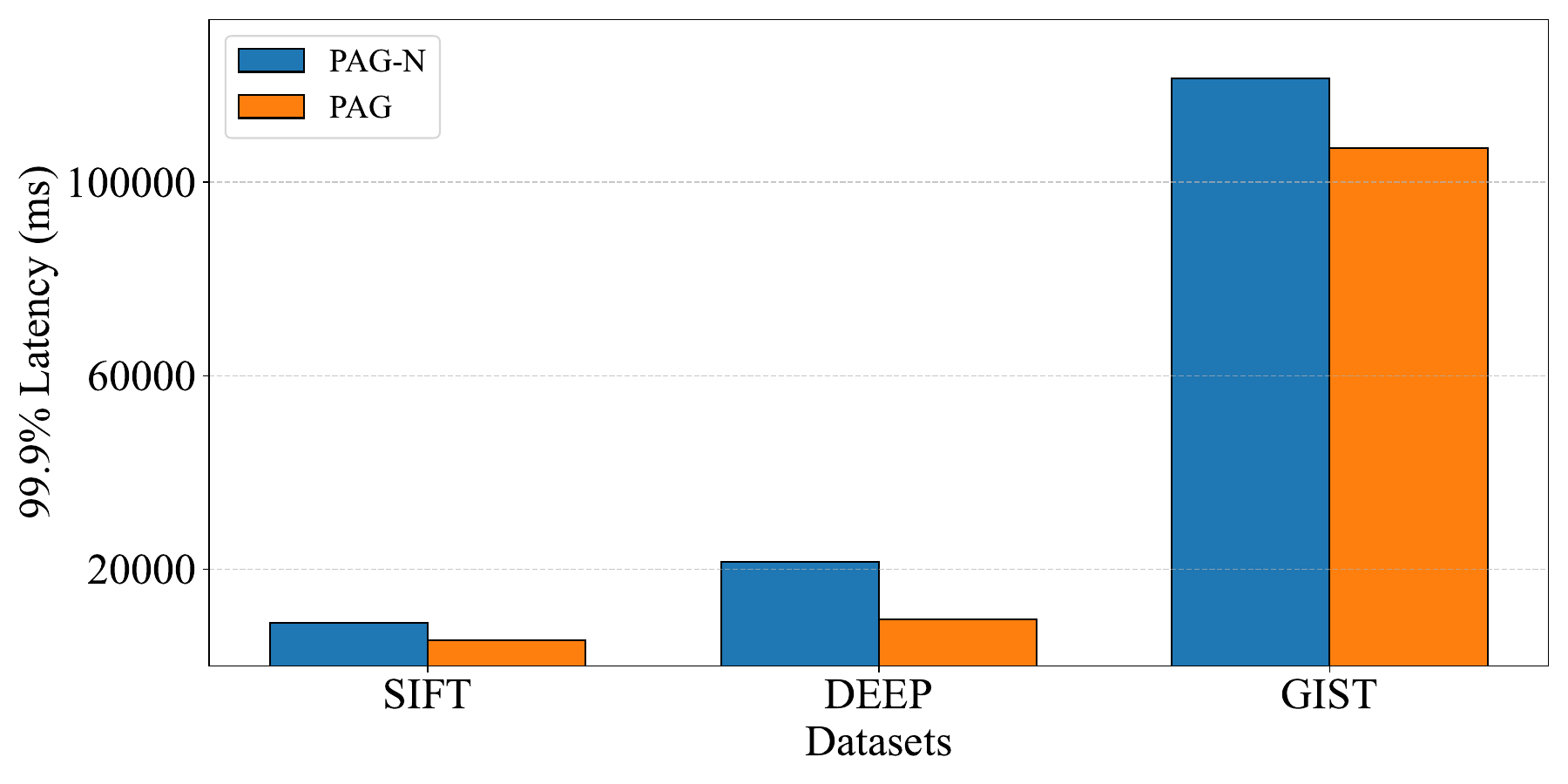}
	\caption{Effect of DRS for Disk-memory hybrid scenario}
	\label{fig:exp_disk_drs}
\end{figure}
\subsection{Ablation Study}
We demonstrate the impact of Dynamic Representation Selection (DRS) on index performance 
using the following configurations: 
(1) PAG without DRS (PAG-N), and (2) PAG with DRS. 
Figure~\ref{fig:exp_disk_drs} shows the 99.9\% query latency for both methods at the same recall rate across different datasets. 
It can be observed that PAG achieves significantly lower 99.9\% query latency compared to PAG-N, 
indicating that DRS effectively balances partition sizes, even though it may introduce some imbalance. 
This effect is particularly pronounced on datasets with significant irregularities, such as DEEP.

\section{Related Work}
\label{sec:related}
\noindent\underline{Graph-based ANNS Methods.}
Graph-based ANNS methods construct a proximity graph (PG) where each data point corresponds to a node
and edges represent proximity or navigability relationships. 
According to the differences of the graph structures,
graph-based methods are divided into several classes.

One of the earliest approaches is the Delaunay Graph (DG) \cite{DBLP:journals/ijpp/LeeS80}, 
which is the dual of the Voronoi Diagram \cite{DBLP:journals/csur/Aurenhammer91} and is proven to be a Monotonic Searchable Network (MSNET). 
However, DG suffers from high node degrees, particularly in high-dimensional space. 
The K-Nearest Neighbor Graph (KNNG) \cite{DBLP:conf/ijcai/HajebiASZ11, DBLP:journals/tcyb/JinZHLCH14} approximates DG by limiting the degree of each node to its $k$ nearest neighbors. 
However, KNNG construction on large-scale datasets is unacceptable 
with a time complexity of $O(n^2)$, and the absence of shortcuts increases search overhead.

To address these limitations, 
methods such as the Relative Neighborhood Graph (RNG) \cite{DBLP:journals/pr/Toussaint80} 
and Sparse Neighborhood Graph (SNG) \cite{DBLP:conf/soda/AryaM93} have been proposed. 
These approaches reduce both construction and search costs by employing edge-pruning strategies 
that approximate KNNG and introduce shortcuts. 
The Diversified Proximity Graph (DPG) \cite{DBLP:journals/tkde/LiZSWLZL20} further improves RNG by ensuring 
that edges are evenly distributed, which optimizes angular coverage. 
However, RNG-based pruning strategy is considered too restrictive for maintaining MSNET property \cite{DBLP:journals/pvldb/FuXWC19}. 
The Monotonic Relative Neighborhood Graph (MRNG) \cite{DBLP:journals/pvldb/FuXWC19}, using a more relaxed pruning strategy, 
retains additional edges to ensure that a monotonic decrease path exists between any two nodes.

Despite their theoretical efficiency, 
constructing RNG or MRNG still incurs high cost.
Practical graph-based ANNS methods such as the Navigating Spreading-out Graph (NSG) \cite{DBLP:journals/pvldb/FuXWC19} 
and Satellite System Graph (SSG) \cite{DBLP:journals/pami/FuWC22} have been developed, leveraging MRNG-based pruning strategy. 
These methods achieve lower construction costs while maintaining competitive recall rates. 
$\tau$-MG \cite{DBLP:journals/pacmmod/PengCCYX23} optimizes the MRNG by fine-tuning the lengths of short and long edges, 
thus improving both Recall@$k$ and exploration efficiency.

Another important class of graph-based ANNS methods is Navigable Small World (NSW) \cite{DBLP:journals/is/MalkovPLK14}, 
known for logarithmic greedy search overhead. 
The Hierarchical NSW (HNSW) \cite{DBLP:journals/pami/MalkovY20} combines RNG-based pruning strategy with hierarchical organization, 
further improving search efficiency.

\noindent\underline{Partition-based ANNS Methods.}
Partition-based ANNS methods divide the data space into subspaces 
where vectors within the same subspace are relatively close. 
Tree-based methods, such as QD-tree \cite{DBLP:conf/sigmod/YangCWGLMLKA20} and KD-tree \cite{DBLP:conf/kdd/RamS19}, 
build a hierarchical structure, where each leaf node represents a subspace 
and internal nodes manage groups of subspaces. 
Hash-based methods, such as Locality-Sensitive Hashing (LSH) \cite{DBLP:conf/vldb/LvJWCL07, DBLP:conf/sigmod/LiYZXCLNC18}, 
map similar points into the same buckets while ensuring balanced bucket sizes. 
An alternative to LSH is Inverted File (IVF) \cite{DBLP:conf/eccv/BaranchukBM18}, 
which clusters data into partitions. 
However, as the size of the dataset grows, boundary points among subspaces become a significant bottleneck, 
often requiring exploration across multiple partitions, which increases the search cost.

\noindent\underline{Compression-based ANNS Methods.}
High-dimensional data presents considerable overhead in distance calculations, 
motivating compression methods designed to reduce data dimensionality. 
Product Quantization (PQ) \cite{DBLP:journals/pami/JegouDS11, DBLP:conf/cvpr/KalantidisA14} is a widely used method 
that clusters vector segments and maps them to the IDs of cluster centers, 
enabling fast approximate distance calculations via pre-computed center-to-center distances. 
Random Projection \cite{DBLP:journals/pacmmod/GaoL23} offers another compression strategy 
by projecting high-dimensional data into lower-dimensional space using a random orthogonal matrix.
These compression methods are often combined with graph-based or partition-based methods, 
enhancing overall efficiency and scalability of ANNS.

\noindent\underline{Hybrid Storage ANNS Methods.}
State-of-the-art methods that support hybrid storage contain two categories: partition-based and graph-based. 
The key idea behind these approaches is storing the compressed vectors in memory while keeping the full vectors in secondary storage. 
Although these methods mainly focus on memory-disk storage, 
they can be easily extended to distributed storage scenarios by replacing the disk with distributed storage.

Graph-based methods are often combined with compression strategies.
DiskANN stores PQ compressed vectors in memory 
while putting the full-precision vectors and the proximity graph on disk.
When a query arrives, DiskANN traverses the graph based on the distance of quantized vectors,
subsequently reranking the candidates according to the distance of the full-precision vectors stored on disk.

Partition-based methods can be naturally viewed as a form of compression.
SPANN uses the Hierarchical Balanced Clustering (HBC) algorithm to generate numerous balanced partitions,
supporting hybrid memory-disk storage by storing the centroids in memory while storing the partition lists on disk.
Vector search using SPANN needs to identify candidate partitions
by calculating the distance between the query and the centroids of the partitions
and then obtains the $K$ nearest neighbors from the candidate partitions via full scan.

\section{Conclusion}
\label{sec:conclusion}
We study graph-cluster hybrid methods for Approximate Nearest Neighbor Search (ANNS) 
in distributed file systems. 
First, we propose DSANN, a novel algorithm that combines graph-based index 
with clustering techniques to efficiently index and search billion-scale vector datasets. 
Additionally, we introduced a concurrent index construction method, 
which significantly reduces the complexity of index building. 
Moreover, we leverage Point Aggregation Graph (PAG) to optimize storage efficiency 
by aggregating similar vectors based on their structural relationships. 
To improve query throughput, we incorporate asynchronous I/O operations 
in the distributed file system. 
Finally, we demonstrate the effectiveness of DSANN through extensive experiments on large-scale datasets, 
showing its superior performance in indexing, storage, and search. 
Experimental results confirm the efficiency and scalability of DSANN in handling high-dimensional vector data in distributed settings.

\bibliographystyle{ieeetr}
\bibliography{add}

\end{document}